\documentclass[nofootinbib,pra,twocolumn,a4paper,groupedaddress,showpacs,floatfix,aps,longbibliography,10pt]{revtex4-1}

\usepackage{graphicx,amsmath,amssymb,amsfonts,dsfont,subfigure,color}

\newcommand{\ket}[1]{\ensuremath{|#1\rangle}}
\newcommand{\mc}[1]{\ensuremath{\mathcal{#1}}}
\newcommand{\bra}[1]{\ensuremath{\langle #1 |}}

\newcommand{\vroW}{\varrho}

\renewcommand{\phi}{\varphi}
\renewcommand{\epsilon}{\varepsilon}
\newcommand{\bR}{\mathbb{R}}

\newcommand{\cL}{\mathcal{L}}

\newcommand{\BK}[1]{ {\left( #1 \right)} }

\newcommand{\curBK}[1]{ {\left\{ #1 \right\}} }

\newcommand{\vecr}{\vec{r}}

\begin{document}

\title{L\'evy statistics of interacting Rydberg gases}

%\title{Level shift and dephasing effects of EIT in an interacting Rydberg gas}

\author{Thibault Vogt,${}^{1,2}$ Jingshan Han,${}^1$ Alexandre Thiery,${}^3$ and Wenhui Li${}^{1,4}$}

\affiliation{${}^1$Centre for Quantum Technologies, National University of Singapore, 3 Science Drive 2, Singapore 117543}
\affiliation{${}^2$MajuLab, CNRS-UNS-NUS-NTU International Joint Research Unit UMI 3654, Singapore 117543}
\affiliation{${}^3$Department of Statistics $\&$ Applied Probability, National University of Singapore, Singapore, 117546}
\affiliation{${}^3$Department of Physics, National University of Singapore, Singapore, 117542}

\pacs{42.50.Gy,32.80.Ee}

% 42.50.Gy Effects of atomic coherence on propagation, absorption, and amplification of light; electromagnetically induced transparency and absorption
% 32.80.Ee Rydberg states

\begin{abstract}
A statistical analysis of the laser excitation of cold and randomly distributed atoms to Rydberg states is developed.  We first demonstrate with a hard ball model that the distribution of energy level shifts in an interacting gas obeys L\'evy statistics, in any dimension $d$ and for any interaction $-C_p/R^p$ under the condition $d/p<1$. 
This result is confirmed with a Monte Carlo rate equations simulation of the actual laser excitation in the particular case $p=6$ and $d=3$. With this finding, we develop a statistical approach for the modeling of probe light transmission through a cold atom gas driven under conditions of electromagnetically induced transparency involving a Rydberg state. The simulated results are in good agreement with experiment.
\end{abstract}

\maketitle

%%%%%%%%%%%%%%%%%%%%%%%%%%%%%
%Introduction
%%%%%%%%%%%%%%%%%%%%%%%%%%%%%
\section{Introduction}

\bigskip
L\'evy statistics applies to the description of a particular class of random walks obeying a `heavy-tailed' probability distribution with power law dependence \cite{bardou2002levy}.
 L\'evy statistics has been studied in great details owing to its importance in numerous scientific areas where such random walks can occur, and where the long tail of L\'evy distributions is essential for the prediction of rare events, be that in chemistry~\cite{ott1990anomalous}, biology~\cite{edwards2007revisiting}, or in economics~\cite{rachev2011financial,ibragimov2015heavy}. In physics, L\'evy statistics  appears often as a result of a complex dynamics, in various transport processes of heat, sound, or light diffusions, in chaotic systems, and in laser cooling where it has broad applications for sub-recoil cooling techniques~\cite{Shlesinger1995levy,barthelemy2008levy,bardou2002levy,ZaburdaevLevy2015}.

It was shown long ago that in a gas the distribution of energy  shifts resulting from the Van der Waals interaction is given by a L\'evy-type distribution, known as van der Waals profile \cite{margenau1935theory}. However, L\'evy statistics has so far not been applied to ensembles of Rydberg atoms, typically interacting via pair-wise long-range potentials. 
Being able to simulate very large systems of interacting Rydberg atoms with simple statistics would be very valuable when the size of the Hilbert space becomes too large to allow for an exact simulation of the quantum dynamics. Moreover, a statistical approach to the simulation of the laser excitation of atoms to Rydberg state would be significantly faster than other approximate computational techniques such as Monte Carlo rate equations (MCRE) sampling \cite{ates2007antiblockade,vogt2007electric,chotia2008kinetic,ates2011electromagnetically,heeg2012hybrid,garttner2012finite,PetrosyanCorrelations2013,GarrtnerSemianalytical2014}.
Most importantly, any approach for simulating the laser excitation must take into account the effect of the interaction blockade \cite{jaksch:00,tong2004local,vogt2006dipole,heidemann2007strongblockade}. The interaction blockade prevents the excitation of pairs of Rydberg atoms with internuclear separations shorter than the so-called blockade distance $D_b$, which may give rise to highly correlated many-body states and tends to favor the excitation of ordered structures \cite{pohl2010dynamical,schwarzkopf2011imaging,schauss2012observation,PetrosyanCorrelations2013}. As a consequence, the dynamics of the excitation of atoms highly depends on the shifts of the individual Rydberg energy levels. For a given atom $i$ and in case of isotropic interactions, this shift due to all the surrounding Rydberg atoms $j$ at $R_{ij}$ distance writes in frequency unit,
\begin{equation}
\nu_i=\sum_j \frac{-C_p}{R_{ij}^p}/h,
\label{freqchangeatomic}
\end{equation}where $h$ is the Planck constant, and $C_p$ is the interaction strength coefficient. The most common interaction is of van der Waals type with $p=6$, while another possibility is the resonant dipole-dipole interaction $p=3$, and in practice both of them can be isotropic for well-chosen experimental conditions. In order to develop a statistical treatment of the laser excitation, it is critical to determine the probability distribution $f \left( \nu \right)$ of energy level shifts $\nu_{i_0}$, at any particular location $i_{0}$ in the system.

The primary purpose of this article is to show that in the case where this location $i_{0}$ is chosen \textit{independently from the position of any excited Rydberg atoms,} $f\left( \nu \right)$ is mostly given by L\'evy stable distributions, and that L\'evy statistics is suitable for the description of the dynamics of an interacting Rydberg gas.
In section \ref{Interacting}, a general statistical analysis of ensembles of interacting particles is reported, valid when the space dimension $d$ and the interaction exponent $p$ satisfy the condition $d/p<1$. In subsection \ref{InteractingIndependent}, we first remind the case of a randomly distributed and interacting gas of point-like particles. This result is interesting for Rydberg atom physics if the bandwidth of the laser excitation is sufficiently large, whereby the blockade distance becomes vanishingly small in comparison to the interatomic distances.
Then in subsection \ref{InteractingHardBall}, we simulate $f \left( \nu \right)$ in the framework of the hard ball model with random distributions of impenetrable interacting particles of radii $D_b/2$, for which the pair-correlation function closely resembles the one obtained considering the excitation dynamics of atoms to Rydberg states \cite{PetrosyanCorrelations2013}. We demonstrate with numerical simulations that the distribution $f\left( \nu \right)$ remains of L\'evy stable type to a very good approximation, and confirm this result analytically at the limit of small $D_b$. In section \ref{LevyLaser}, the actual laser excitation of a cold gas to Rydberg state is studied specifically for $(p,d)=(6,3)$ in a three-level ladder excitation scheme. We use the MCRE approach in order to simulate the two-photon excitation with near-resonant lasers, and confirm the distribution $f\left( \nu \right)$ yielded by the hard ball model.
Finally in section \ref{RydbergEIT}, we develop a model of the probe light transmission through an atomic gas driven in configuration of electromagnetically induced transparency involving a Rydberg state (Rydberg EIT), with statistical modeling of the interactions based on L\'evy distributions. The simulated transmission spectra match closely those obtained directly by the MCRE approach, and are in good agreement with experimental data acquired as in reference \cite{han2016spectral}.

\section{Levy statistics of interacting particles}
\label{Interacting}
\subsection{L\'evy statistics of independent point-like particles}
\label{InteractingIndependent}

This subsection provides a reminder on the distributions of energy level shifts $f \left( \nu \right)$ for randomly distributed point-like particles interacting via $-C_p/R^p$ pair-wise potentials. We assume $C_p<0$ for convenience, although all the calculations presented in this section are also valid for $C_p>0$, as long as the change of sign is properly taken into account in all the expressions of the distribution functions.

As demonstrated in Appendix \ref{Tmodel} for any isotropic interaction of $-C_p/R^p$ type and any dimension $d$, $d/p<1$, the distribution $f\left( \nu \right)$ is equal to the L\'evy stable distribution $f_{\alpha,\lambda_0}\left( \nu \right)$, whose Laplace transform takes the simple form
\begin{equation}
\cL_{\alpha,\lambda_0}(t)= \exp\left(-\lambda_0 \, t^{\alpha}\right).
\end{equation}
The two parameters in this equation are $\alpha=d/p$, and 
\begin{equation}
\lambda_{0} =\frac{\Gamma(1-d/p)  \, S_d }{d}  \left(-C_p/h\right)^{\frac{d}{p}} n_r, 
\end{equation}
where $n_r$ is the density of interacting particles, $\Gamma$ stands for the Gamma function and $S_d$ is the unit surface in $d$-dimension, namely $S_1=2$, $S_2=2 \pi$, or $S_3=4 \pi$.

In the case $(p,d)=(6,3)$, the distribution of shifts $f\left( \nu \right)$ is therefore given by the following distribution, obtained long ago and known as L\'evy distribution or van der Waals profile \cite{margenau1935theory}
\begin{equation}
f_{1/2,\lambda_0}(\nu)=\frac{\lambda_0}{2 \sqrt{\pi}} \, \nu^{-3/2} \, e^{\frac{-\lambda_0^2}{4 \, \nu}},
\label{Levy0}
\end{equation}
where $\lambda_0$ simplifies in this case to
\begin{equation}
\lambda_0=\frac{4 \pi^{3/2}}{3}\sqrt{-C_6/h} \, n_{r}.
\end{equation}
This distribution is highly peaked at $\nu_{max}=\lambda_0^2 /6$ and possesses a very slow decreasing tail which is mostly due to the interaction with the nearest neighbor particle.  
Simple analytical expressions of the distribution functions exist for other choices of the parameters $p$ and $d$ as well, especially for $d/p=1/3$ and $d/p=2/3$ as given in Appendix \ref{Tmodel}. 

\subsection{L\'evy statistics in the hard ball model}
\label{InteractingHardBall}

The outstanding question addressed in this subsection is how the distributions of energy level shifts $f \left( \nu  \right)$ obtained in the previous subsection are modified when the particles become impenetrable. For this purpose, we consider a system composed of interacting hard balls of radii $D_b/2$, and perform numerical simulations by Markov Chain Monte Carlo (MCMC) computations of hard balls, using an algorithm similar to the so-called Markov-chain hard-sphere algorithm of Ref.~\cite{krauth2006statistical}. 
Remarkably, we find that the distribution of energy shifts $f \left( \nu  \right)$ remains of L\'evy stable type to a very good approximation, up to
\begin{figure}[ptb]
\includegraphics[width=8cm]{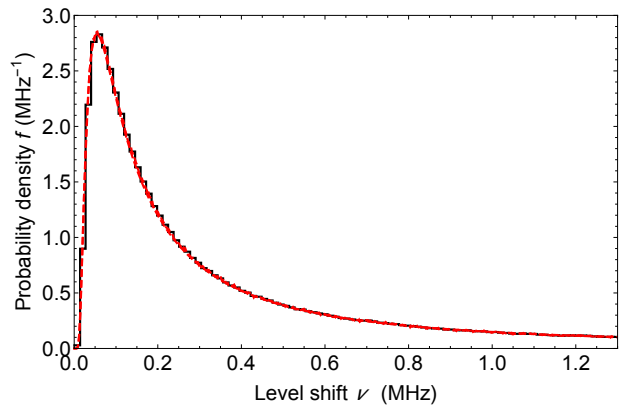}%
\caption {(color online) Comparison between an histogram of the distribution of energy level shifts $f \left( \nu \right)$ computed by Markov Chain Monte-Carlo (MCMC) simulation of hard balls (black solid line), and the L\'evy distribution function $f_{\frac {1}{2},\lambda_0 \times \left(1+ 0.26 D_b^3 /R_3^3\right)}(\nu)$ with $\lambda_0=\frac{4 \pi^{3/2}}{3} n_{r} \sqrt{-C_6/h}$ where $C_6/h=-1 \, \textrm{MHz} \cdot \mu\textrm{m}^6$ and $n_{r}=7.07 \times 10^{10} \, \textrm{m}^{-3}$ (red dashed line). The system is composed of a cube of $33.8\ \mu$m edge length, containing hard balls of diameter $D_b=1.05\ \mu$m with Wigner-Seitz radius $R_3=\left( \frac{1}{4/3 \pi n_r} \right)^{1/3}=D_b/0.7=1.5\ \mu$m, and periodic boundary conditions are assumed. The histogram of $f \left( \nu \right)$ is realized with the computed data of the total shift at the center of the cube, $\sum \frac{-C_6/h}{R^6}$, accumulated over 8000 different trajectories, with a different initial state and 200000 Markov chain steps for each trajectory. \label{ComparisonLevyHS}}
\end{figure}
$D_b \sim R_d$, where $R_d$ is the Wigner-Seitz radius 
\begin{equation}
R_d \equiv \left(\frac{d }{S_d n_r }\right)^{1/d}.
\end{equation}
As illustrated in Fig.~\ref{ComparisonLevyHS} for the particular case of $(p,d)=(6,3)$, the numerical simulations are very well fitted by $f_{\alpha,\lambda}$ L\'evy stable distributions, where $\lambda$ is used as a free fit parameter, and $\alpha=d/p$. The $\lambda$ parameter extracted from fitting of the simulated distributions verifies the relation
\begin{equation}
\lambda=\lambda_0 \left(1+g\left( D_b/R_d \right) \right),
\end{equation}
where $g\left( D_b/R_d \right)$ is a correction term which can be written in the form of a polynomial of the variable $D_b/R_d$. At the lowest order, this term writes
 \begin{equation}
 g\left( D_b/R_d \right) \approx \alpha \, \frac{D_b^d}{R_d^d},
 \end{equation}
 where $\alpha \sim 0.3$ when $(p,d)=(6,3)$, as shown in Fig.~\ref{CorrectionCoefHS}.
The asymptotic form of the distribution $f \left( \nu \right)$ can be recovered analytically in the limit of small $D_b/R_d$ as shown in Appendix \ref{Tmodel}.
We find that for any interaction exponent $p$ and dimension $d$, $d/p<1$, the distribution of energy shifts $f\left( \nu \right)$ is given by a L\'evy stable distribution $f_{\frac{d}{p},\lambda}\left( \nu \right)$ of parameter
\begin{equation}
\lambda=\lambda_{0} \, \BK{ 1 + \epsilon \, n_r \, V(D_b) },
\end{equation}
with
\begin{equation}
\epsilon=\frac{2-2^{d/p}}{2},
\label{AnaResult}
\end{equation}
and where $V(D_b)=S_d D_b^d /d $ is the $d$-dimensional volume of a ball of radius $D_b$.
This analytical approximation is in good agreement with Monte-Carlo simulations at small $D_b/R_d$ ratios, as illustrated in Fig.~\ref{CorrectionCoefHS} for the case $(p,d)=(6,3)$ where $\alpha \approx \epsilon$. The higher order terms at large $D_b/R_d$ ratios are still under investigation, but they seem to account only for a small additionnal correction, as illustrated in Fig.~\ref{CorrectionCoefHS} as well.  
The lowest order dependence on the exclusion volume $V(D_b)$ is not surprising as it appears also in the derivation of the second Virial coefficient of the equation of state of a van der Waals gas \cite{krauth2006statistical}. The shift of the distribution peak to a higher frequency, from $\nu_{max}=\lambda_0^2/6$ to $\lambda^2/6$, is a manifestation of the ordering at higher $D_b$, as correlations tend to decrease the probability for a region locally empty of hard balls. The long tail of the distribution, which varies as $\nu^{-3/2}$ for large shifts in Fig. \ref{ComparisonLevyHS}, is due mostly to the presence of the nearest neighbor. Indeed, since the level shift is calculated at a particular position chosen independently from the positions of the hard balls, the tail of the distribution still varies as in the case of point-like particles.
\begin{figure}[hptb]
\includegraphics[width=8cm]{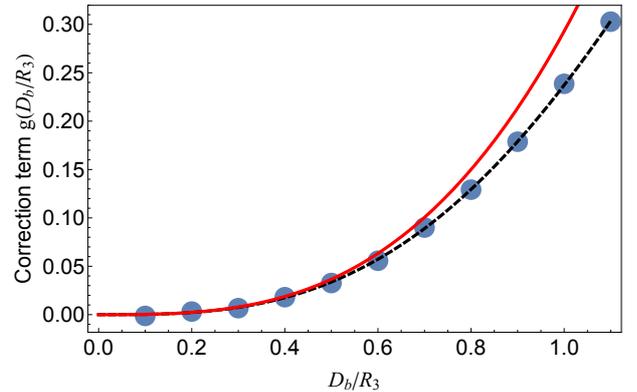}%
\caption {(color online) Correction term $g(x)$, as a function of $x=D_b/R_3$. The analytical approximate curve (red solid line) is given by $g(x)=\epsilon \, x^3$ with $ \epsilon= \frac{2 - \sqrt{2}}{2}$. The data points (blue circles) are extracted from fitting of the simulated energetic distributions $f \left( \nu \right)$ to a L\'evy distribution $f_{\frac{1}{2},\lambda_0 \times \left(1+ g(x)\right)}(\nu)$ of parameter $\lambda_0 \times \left(1+ g(x)\right)$. The distributions $f \left( \nu \right)$ are simulated with MCMC computations as in Fig. \ref{ComparisonLevyHS}, at fixed particles density, and with various $D_b$. The fit of the data points to a model curve of the form $g(x)=\alpha \, x^3+\beta \, x^5 $ gives $\alpha=0.281 \approx \epsilon$ and $\beta=-0.044$  (black dashed line). \label{CorrectionCoefHS}}
\end{figure}

\section{L\'evy statistics of laser excited Rydberg atoms}
\label{LevyLaser}

As shown in the previous section, the distribution of energy shifts $f \left( \nu \right)$ is given in excellent approximation by a L\'evy stable distribution for a system of hard balls interacting via $-C_p/R^p$ long-range potentials, in any dimension $d$ such that $d/p<1$. 
While restricting the discussion in this section to the case $(p,d)=(6,3)$, we show that the steady-state distribution of shifts $f  \left( \nu \right)$ is also a L\'evy distribution in a system of laser excited Rydberg atoms, where the Rydberg atoms interact via $-C_6/R^6$ long-range potentials. In this system, the dipole blockade prevents the excitation of pairs of Rydberg atoms with interatomic distances $r$ shorter than the blockade distance $D_b$. As a result of the blockade, the radial pair-correlation function $g_2(r)$, characterizing the probability of having one Rydberg atom at a distance $r$ from another Rydberg atom, is very similar to that obtained for a system of hard balls of radii $D_b/2$, but has a smoother transition at the distance $r\sim D_b$. We will study in this section the analogies between the distribution of shifts in these two systems, but also the slight differences resulting from the dynamics of the laser excitation.

\begin{figure}[hptb]
\includegraphics[width=4cm]{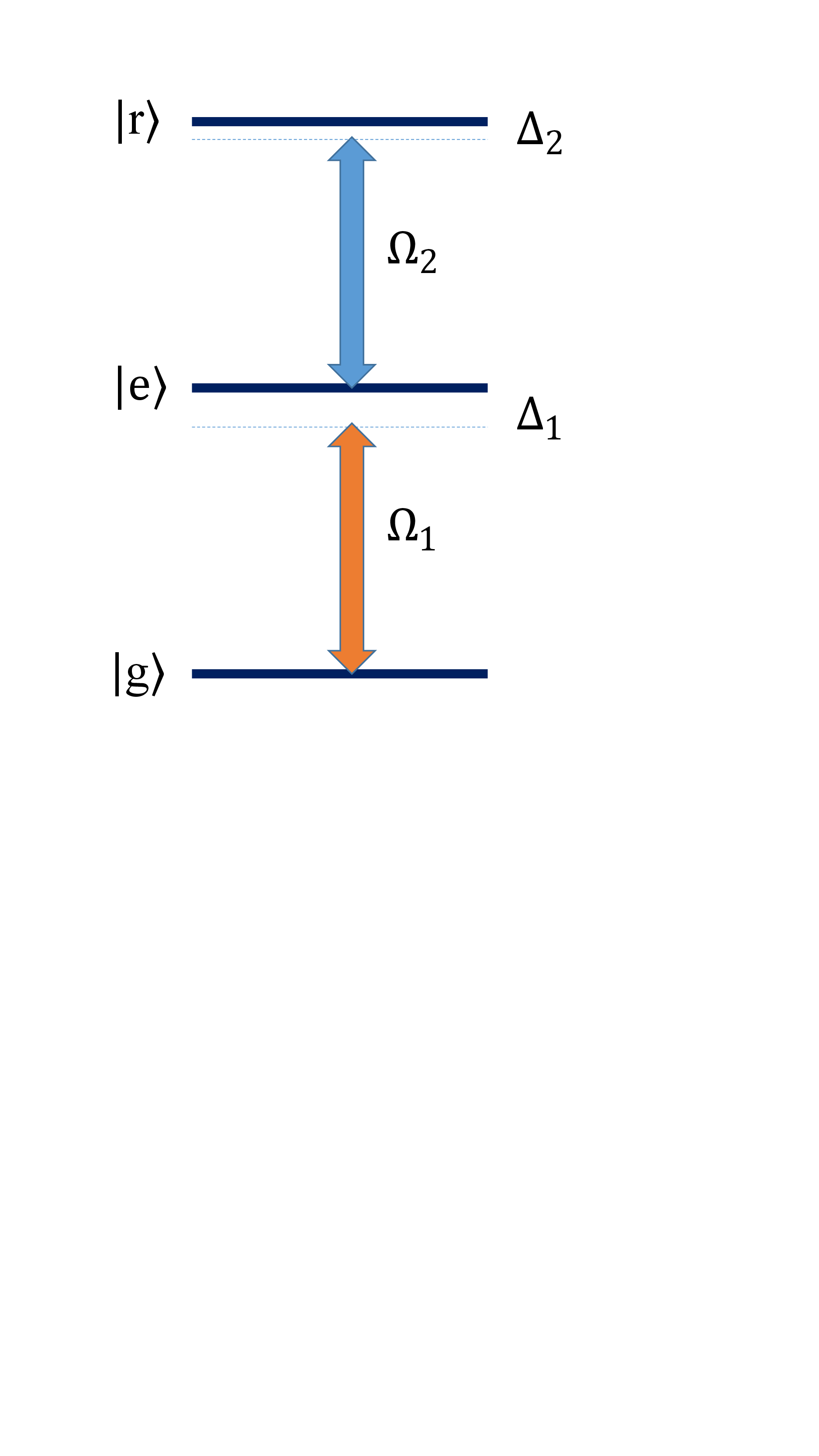}%
\caption {(color online) Scheme for laser excitation of ground state atoms to a Rydberg state as used in the Monte Carlo rate equations (MCRE) computation. Two laser beams with Rabi frequencies $\Omega_1$ and $\Omega_2$ are detuned $\Delta_1$ and $\Delta_2$ from the resonance of the $|g\rangle \rightarrow |e\rangle$  and $|e\rangle \rightarrow |r\rangle$ transitions, respectively. \label{SetupRydEIT}}
\end{figure}

A common method for producing Rydberg atoms, shown in Fig.~\ref{SetupRydEIT}, consists of a ladder scheme of excitation with two lasers (1) and (2), coupling the $|g\rangle \leftrightarrow |e\rangle$ and $|e\rangle \leftrightarrow |r\rangle$ transitions, respectively, where $|g\rangle$,  $|e\rangle$  and $|r\rangle$ are three atomic levels, ground, intermediate and Rydberg, respectively. Without loss of generality, we only study the case of $^{87}\mathrm{Rb}$ atoms, with the three energy levels $|g \rangle =|5s_{1/2} \rangle$, $|e \rangle =|5p_{3/2} \rangle$ whose decay rate is $\Gamma_e /2 \pi=6.067\ $MHz, and $|r \rangle =|\text{n} s \rangle$, where the principal quantum number $\text{n}$ can be varied for tuning the $C_6$ parameter~\cite{singer2005long}. Furthermore, we consider an homogeneous gas of ultra-cold atoms at atomic densities $n_{at} \leq 5\times 10^{11}\ \mathrm{m}^{-3}$, and the laser fields $(1)$ and $(2)$ are assumed to be classical.

We calculate the distribution of shifts $f \left( \nu \right)$ at the positions of atoms $i$ in states $|g \rangle$ and $|e \rangle$, which is of most relevance when considering the laser excitation to Rydberg state of such atoms $i$ in the midst of atoms $j$ already excited to Rydberg state.
The steady-state distribution of energy level shifts is computed using the MCRE approach, and according to the method described in reference \cite{GarrtnerSemianalytical2014} and summarized in Appendix \ref{LevyModel}. The MCRE approach is a well-known method for computing the excitation dynamics in such a system, which can provide approximate results when the system quickly relaxes to the steady state as is the case for near-resonant excitation \cite{ates2011electromagnetically, garttner2013nonlinear}.
\begin{figure}[hptb]
\includegraphics[width=8cm]{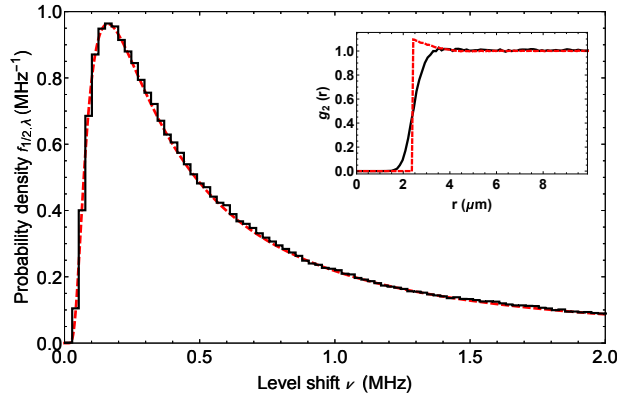}%
\caption {(color online)  Comparison between an histogram of the distribution of energy shifts $f \left( \nu \right)$ computed using the MCRE approach (black solid line) and the L\'evy distribution $f_{\frac{1}{2},\lambda_0 \times \left(1+0.31 \, D_b^3 /R_3^3 \right)}$ (red dashed line). The system is composed of a cube containing 1000 atoms at a ground state density of $n_{at}=10^{11}\ \mathrm{m}^{-3}$ and the computation is carried out with periodic boundary conditions.  $f \left( \nu \right)$ corresponds here to the distribution for an equilibrium excitation of $^{87}\mathrm{Rb}$ atoms, with $\Omega_1/ 2 \pi=1.6$~MHz, $\Omega_2 /2 \pi=5.6$~MHz, $\Delta_1=\Delta_2=0$, Rydberg state $|r \rangle=|38s \rangle$, and decay rate of atomic coherences $\gamma_{ge} \approx \Gamma_e/2$ and $\gamma_{gr}/ 2 \pi= 0.1$~MHz. The Wigner-Seitz radius is equal to $R_3=\left( \frac{1}{4/3 \pi n_r} \right)^{1/3}\sim \, D_b/0.67$, where $D_b=2.4\ \mu$m.  The histogram is plotted with the Rydberg energy level shifts of all the atoms excluded the ones that are in the Rydberg state, computed over 23 different trajectories, with nearly 1300 Monte Carlo time steps each. The inset shows the radial pair-correlation function $g_2 (r)$ in this system (black solid line), compared to that of the ideal system of hard balls of diameter $D_b$ obtained considering the same density $n_r$ (red dashed line). Both pair-correlation functions are computed with Monte-Carlo simulations.  \label{ComparisonLevyMC}}
\end{figure}
As can be seen on Fig.~\ref{ComparisonLevyMC} for the conditions indicated in the figure caption, the distribution of energy shifts $f$ closely resembles a L\'evy distribution 
\begin{equation}
f(\nu) \approx f_{\frac{1}{2},\lambda_0\left(1+ \alpha \frac{4}{3} \pi n_r D_b^3 \right)}(\nu),
\label{approxDist}
\end{equation}
where $\alpha\sim \epsilon \sim 0.3$, while $n_r=f_r \, n_{at}$ is the density of Rydberg atoms proportional to the fraction of atoms in the Rydberg state $f_r$ and the atomic density $n_{at}$. Moreover, in Eq. \eqref{approxDist}, $D_b$ is the blockade distance, which we define rigorously as the distance that satisfies
\begin{equation}
\vroW_{rr}\left(\frac{C_6/h}{D_b^6}\right)=\frac{\vroW_{rr}(0)}{2},
\end{equation}
where $\vroW_{rr}\left(-\nu \right)$ is the single-atom Rydberg population calculated for a given Rydberg energy level shift $h \nu$, as obtained from the solution $\vroW$ of the single-atom master equation reminded in Appendix \ref{LevyModel}. The profound analogy between the hard ball model and the real physical system becomes obvious when looking at the radial pair-correlation function in the inset of Fig.~\ref{ComparisonLevyMC}, and this is the main reason why the distributions of energy level shifts in both systems look alike.

Next, we come to the range of validity of the approximation of the distribution of shifts by a L\'evy distribution. In the hard balls model, $f \left( \nu \right)$ is given in excellent approximation by a L\'evy stable distribution, up to densities of the order $n_r \sim n_b \,= \, 3/(4 \pi D_b^3)$, which corresponds to $R_3$ down to $R_3 \sim D_b$. Actually, this condition corresponds to a density of interacting particles that is already relatively high, knowing the densest possible packing of hard balls is obtained for the honeycomb lattice where $R_3 = D_b/\left(4 \pi \sqrt{2}/3\right)^{1/3} \sim D_b/1.8$. In laser excitation of Rydberg atoms, at most one Rydberg atom can be excited per blockade volume $V\left( D_b\right)=4 \pi D_b^3/3$ as given by the super-atom model, which translates into the condition $R_3 \geq D_b$~\cite{PetrosyanCorrelations2013,GarrtnerSemianalytical2014}. Densities slightly larger are achievable in our system, as we obtained with the MCRE computation, and in agreement with recent simulations~\cite{GarrtnerSemianalytical2014}. However we can easily verify that our system does not draw near close packing where the distribution of shifts would be fundamentally altered.

In order to verify more systematically the approximation of the distribution of shifts by a L\'evy distribution, we have simultaneously varied the total atomic density $n_{at}$ in the range of 0.5 to $2 \times 10^{11}\ \mathrm{cm}^{-3}$ and $C_6/h$ in the range of  $-0.8 \, \textrm{MHz} \cdot \mu \textrm{m}^6$ to $-2500 \, \textrm{MHz} \cdot \mu \textrm{m}^6$, for different choices of $(\Delta_1,\Omega_1,\Omega_2)$, while keeping $\Delta_2=0$. We have also varied simultaneously the detuning $\Delta_1 /2 \pi$ from -3 to +3 MHz, the Rabi frequency $\Omega_1 /2 \pi$  from 1.5 to 6 MHz, and the Rabi frequency $\Omega_2 /2 \pi$  from 1 to 6 MHz, while keeping $\Delta_2=0$. We find that the distributions $f(\nu)$ are well approximated by L\'evy distributions $f_{1/2,\lambda}(\nu)$  of parameter
 \begin{equation}
 \lambda \approx \lambda_0 \left(1+ \alpha \frac{ D_b^3/R_3^3 }{\rho_{gg}^{(S)} +\rho_{rr}^{(S)} D_b^3/R_3^3}\right) 
 \label{approxlambda}
 \end{equation}
where $\rho_{gg}^{(S)}$ and $\rho_{rr}^{(S)}$ are the steady state population solutions of the single-atom master equation, and $\alpha$ is nearly equal to $\epsilon$. 
To characterize the quality of the fit by a L\'evy distribution, we calculate the total variation distance 
\begin{equation}
\mc{D}=\int_0^{+\infty} d\nu \, |f(\nu)-f_{1/2,\lambda}(\nu)|/2,
\end{equation}
and find $\mc{D}\, < \, 0.07$ for all the tested conditions. Since $D_b/R_3$ reaches as much as 1.15 in the simulated set of data, we can conclude with reasonable certainty that the L\'evy distribution constitutes a good approximation for our system.
In condition of Rydberg EIT, we have $\Omega_1 \ll \Omega_2$, hence $\rho_{rr}^{(S)} \ll \rho_{gg}^{(S)}$, which means $\lambda$ indeed has the form shown in Eq.~\eqref{approxDist}. It is interesting to note that at the opposite limit, $\Omega_1 \gg \Omega_2$, we have $\rho_{rr}^{(S)} \gg \rho_{gg}^{(S)}$, and $\lambda\approx \lambda_0 (1+\alpha)$.
 $\alpha$ is nearly constant in the range of parameters covered by our simulations except for a small variation from 0.2 to 0.4 versus detuning $\Delta_1$. This variation may be due to the pair correlation function enhancement on the blue two-photon detuning $\Delta_1+\Delta_2 >0$ \cite{schempp2014full,valado2015experimental}, and is still under investigation.

An important conclusion drawn from this calculation is that the distribution $f \left( \nu \right)$ of level shifts in this system closely matches a simple L\'evy distribution. The effect of the dipole blockade is accounted almost entirely in a single coefficient $\alpha  \sim \epsilon$ given by Eq. \eqref{AnaResult}, although we have identified an important correction to the expected hard ball model in the limit of $\Omega_1 \gg \Omega_2$. 
Generally, the distribution of shifts directly reflects on the resonance profiles that can be measured by spectroscopy, hence the simulation results reported here could be easily tested. 
%For example, the correlation coefficient $\alpha$ in Eq. \eqref{approxDist} may be obtained by measuring the excitation fraction $f_r$ versus $\Delta_2$ (or quite similarly versus $\Delta_1$), with the assumption that $f_r$ satisfies the following equation 
%\begin{equation}
%f_r \left( \Delta_2 /2 \pi \right) =\int_0^{+\infty}\varrho_{rr}\left(\Delta_2/2\pi-\nu \right) f \left( \nu \right)  \text{d}\nu,
%\label{frLevy}
%\end{equation}
%where $f$ depends also on $f_r$. 
In principle, it is possible to fully probe certain distribution of shifts by measuring the fraction $f'_r$ of atoms excited to a Rydberg state $|\text{n}' s\rangle$ versus frequency detuning of an excitation laser, $\Delta \nu$, and in presence of atoms already excited to another state $|\text{n}s\rangle$
\begin{equation}
f'_r \left( \Delta \nu \right) =\int_0^{+\infty}P'_r\left(\Delta \nu- \nu \right) f \left( \nu \right)  \text{d}\nu  \propto f \left( \Delta \nu \right),
\end{equation}
where the single-atom probability of excitation to $|\text{n}' s\rangle$, $P'_r$, is assumed to be a very narrow function compared to the distribution of shifts $f$, as realized for example in conditions of weak driving. 

\section{L\'evy statistics applied to Rydberg EIT}
\label{RydbergEIT}

Many experimental and theoretical studies have investigated the effects of dipolar interactions on Rydberg EIT-like spectra \cite{pritchard2010cooperative,weimer2008quantum,petrosyan2011electromagnetically,ates2011electromagnetically,sevinccli2011nonlocal,sevinccli2011quantum,heeg2012hybrid,garttner2012finite,PetrosyanCorrelations2013,GarrtnerSemianalytical2014}.
Recently, it has been demonstrated that the energy level shifts in an interacting Rydberg gas are responsible for a readily observable spectral shift of the EIT transparency window, in addition to the strong dephasing that was observed in previous experiments \cite{desalvo2016rydberg,han2016spectral}. 
The purpose of this section is to show that a light propagation model based on L\'evy statistics can capture most of the features observed experimentally.

Rydberg EIT consists in driving the two transitions of Fig. \ref{SetupRydEIT}, with weak probe (1) and strong coupling (2) near-resonant lasers. For clarity, subscripts  1 and 2 of the previous section will be replaced by subscripts $p$ and $c$, respectively.
The propagation of the probe light in an EIT medium is generally well described within the framework of coupled Maxwell-Bloch equations. In the following analysis, we only consider the case $\Delta_c=0$. Moreover, the kinetics of the atoms is neglected, while steady state equilibrium of the excitation is still assumed, as it is reached on a timescale much shorter than the duration of the typical experiment. The system is assumed to be sufficiently large and homogeneous to apply the local density approximation. We consider the field as classical and neglect photon-photon correlations, while light modulation 
instabilities are also neglected \cite{sevinccli2011nonlocal}, which is consistent with the atomic densities and light intensities used in Ref.~\cite{han2016spectral}. Finally, we will neglect any lensing or diffraction effects~\cite{han2015lensing}.
Consequently, Maxwell's equations reduce after applying the slow envelope approximation to the following one-dimensional differential equation for the probe field \cite{han2015lensing}
\begin{equation}
\partial_{z} \mc{E}_{p}\left( z \right)=   \frac {i k} {2} \chi \left( z \right) \mc{E}_{p}\left( z \right),
\label{maxSS}
\end{equation}
where $\chi \left( z \right)$ is the atomic susceptibility  along the probe beam propagation direction $z$, and $\mc{E}_{p}\left( z \right)$ is the amplitude of the probe field. Then, for a dilute gas, the probe intensity $I_p$ is simply obtained from the differential equation 
\begin{equation}
\partial_{z} I_{p}\approx  -k \mathrm{Im} \left[ \chi \right] I_p.
\label{equaIp}
\end{equation}
In principle, the calculation of the susceptibility involves the solution of the full master equation of the many-body system. 
In our system, the susceptibility $\chi$ can be approximated as a local spatial average over the optical responses of the different atoms. 
In a simple manner, the susceptibility can be written as the following average over interaction configurations \footnote{A simple relation between $\chi$, the Rydberg fraction $f_r$, the two-level atom susceptibility $\chi_{2L}$, and the three-level atom susceptibility $\chi_0$ was found from Monte-Carlo simulations \cite{ates2011electromagnetically,GarrtnerSemianalytical2014}, which is however verified only for zero probe detuning.}
\begin{equation}
\chi =\left(1 -f_r \right) \int_0^{+\infty} \chi_0 (\nu) f(\nu) d\nu,
\label{chiExact}
\end{equation}
where $ f(\nu)$ is the distribution function of energy level shifts due to the Van der Waals interaction, $f_r$ is the averaged Rydberg population fraction, and $\chi_0(\nu)$ is the single-atom 3-level susceptibility calculated at a given Rydberg energy level shift $ h \, \nu$ or equivalently for an effective coupling laser detuning $\Delta_c=-2 \pi \nu$ (cf. Appendix \ref{LevyModel} for a more detailed definition of $\chi_0$). 
As $ f(\nu)$ is given in a very good approximation by the L\'evy distribution $f_{\frac{1}{2},\lambda_0\left(1+ \epsilon \frac{4}{3} \pi n_{at} f_r D_b^3 \right)}$ with $\lambda_0= \frac{4 \pi^{3/2}}{3} \sqrt{-C_6/h} \, n_{at} f_r$, only the knowledge of $f_r$ is needed in Eq. \eqref{chiExact} in order to calculate the susceptibility. 
According to the super-atom model \cite{petrosyan2011electromagnetically,PetrosyanCorrelations2013,GarrtnerSemianalytical2014}, a reasonable analytical approximation for $f_r$ is obtained from the following equation
\begin{equation}
f_r=\frac{f_0}{1-f_0+f_0 n_{at} V_B},
\label{Rydfrac2}
\end{equation}
where $V_B=\frac{4}{3} \pi D_b^3$ is the volume of the blockade sphere of radius $D_b$, and $f_0 \equiv \vroW_{rr}(\nu=0)$ is the Rydberg population fraction in absence of interactions.
\begin{figure}[hptb!]
\centering
\includegraphics[width=8.5cm]{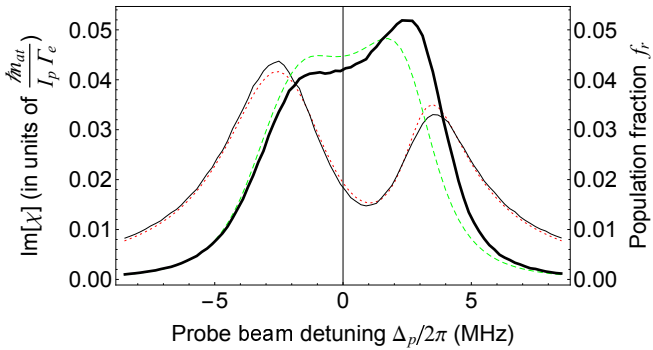}%
\caption {(color online) Imaginary part of the susceptibility and $|38s\rangle$ Rydberg state population fraction $f_r$ for an ensemble of interacting atoms at a density of $n_{at}=1.25\times 10^{11}\ \mathrm{cm}^{-3}$, driven in configuration of EIT with probe field Rabi frequency $\Omega_p / 2 \pi=1.45$~MHz and coupling field Rabi frequency $\Omega_c / 2 \pi=5.6$~MHz. The plotted curves are as follows: $\mathrm{Im} \left[\chi \right]$ obtained from Eq. \eqref{chiExact}, where the Rydberg fraction $f_r$ is calculated with Eq. \eqref{Rydfrac2} (red dotted line); $\mathrm{Im} \left[\chi_{MC} \right]$ computed with the MCRE simulation over 800 atoms in a square box (black thin solid line); the population fraction in the Rydberg state $f_r$ obtained directly with the MCRE simulation (black thick solid line); and the approximate Rydberg population fraction $f_r$ as given by Eq. \eqref{Rydfrac2} (green dashed line).   \label{AppendixFig}}
\end{figure}

The imaginary part of the susceptibility as calculated with Eqs. \eqref{chiExact} and \eqref{Rydfrac2}, $\mathrm{Im} \left[\chi \right]$, may be directly compared to that obtained from MCRE simulations 
\begin{equation}
\mathrm{Im} \left[\chi_{MC} \right]=\frac{\hbar n_{at}}{I_p \Gamma_e} f_e,
\end{equation}
where $f_e$ is the average intermediate state population fraction, and $n_{at}$ is the atomic density.
Although none of the two approaches yield exact solutions for the susceptibility, they lead to very similar results, as shown in Fig. \ref{AppendixFig}, except for a small discrepancy near the Autler-Townes resonances. There is actually a larger mismatch on the blue detuning side of the $|g\rangle \rightarrow |e \rangle $ transition when we compare the Rydberg population fractions given by Eq. \eqref{Rydfrac2} and that yielded by the MCRE simulation, as shown in Fig. \ref{AppendixFig}.
The susceptibility, proportional to the intermediate state population fraction, is however not very sensitive to the Rydberg population fraction at such detuning, where the rate of single-photon excitation is high. This explains the good agreement between $\mathrm{Im} \left[\chi \right]$ and $\mathrm{Im} \left[\chi_{MC} \right]$ even for probe detuning on the blue side of the $|g\rangle \rightarrow |e \rangle $ transition.

\begin{figure}[hptb]
\includegraphics[width=8.5cm]{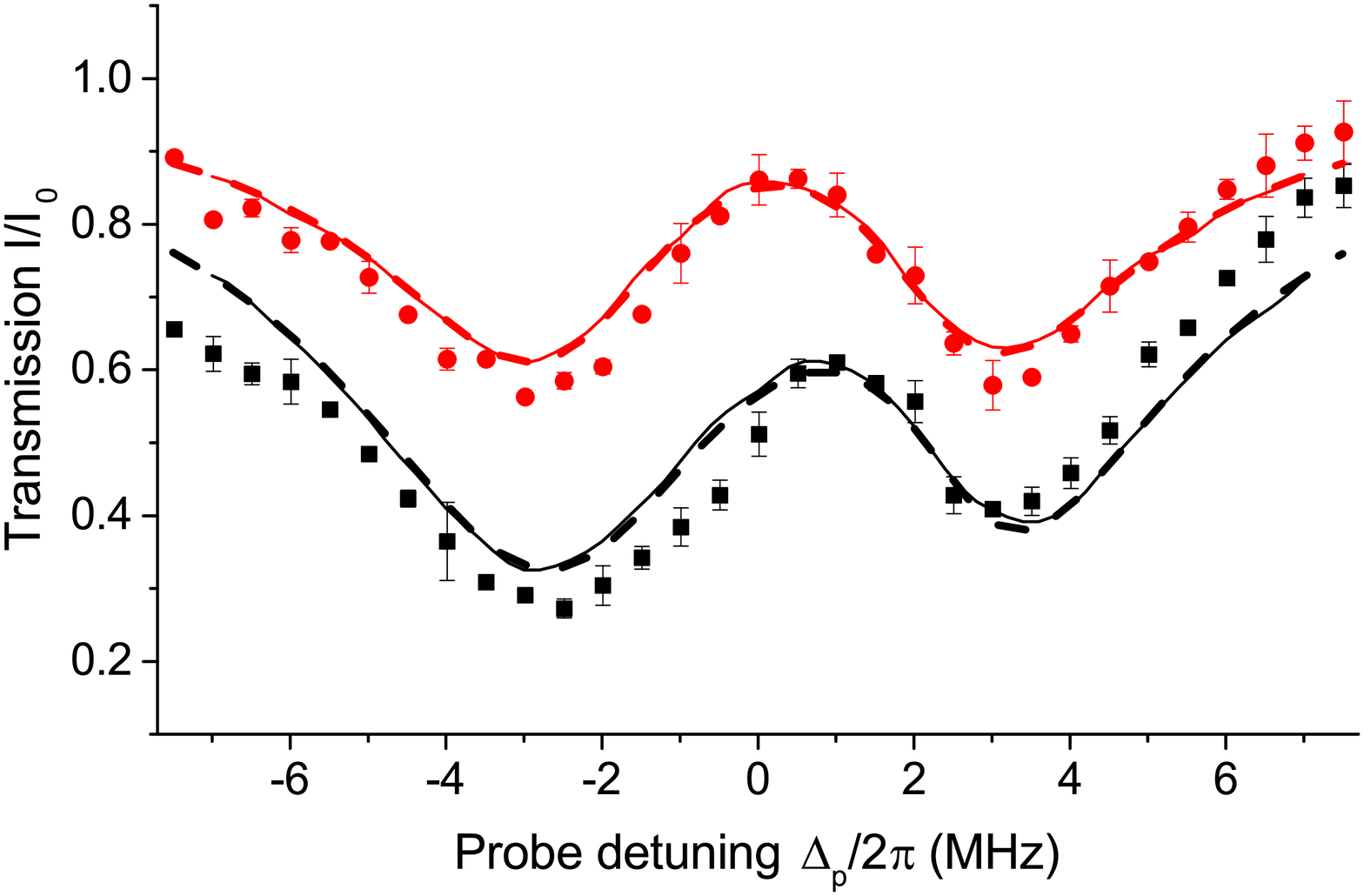}%
\caption {
(color online) Comparison between the probe transmission spectra acquired in an experiment similar to that of reference \cite{han2016spectral} and generated by theoretical models. The experimental spectra are acquired with a coupling field Rabi frequency of $\Omega_{c}/2\pi = 5.6\ \mathrm{MHz}$, an input probe field Rabi frequency of $\Omega_{p0}/2\pi = 1.45\ \mathrm{MHz}$, the Rydberg state $|r \rangle=|38s \rangle$, and in the case of two different Gaussian atomic clouds. The first one is of $1/e^2$ radius $w_z = 51\ \mu m$ with peak atomic density, $n_0 = 0.38\times10^{11}\ \mathrm{cm}^{-3}$ ($\bullet$), while the second one is of $1/e^2$ radius $w_z =24\ \mu m$ with peak atomic density $n_0 =1.76\times10^{11}\ \mathrm{cm}^{-3}$ ($\blacksquare$).  The dashed lines are generated by the L\'evy statistics model with the corresponding experimental parameters as inputs while the solid lines are obtained using the MCRE approach (see text).\label{Figure2Levy}}
\end{figure}

We have seen so far how L\'evy statistics may be used successfully for describing Rydberg EIT by comparison to MCRE computations. In order to fully validate the approach, we provide in the following paragraph a direct comparison between the simulation and some experimental data acquired in conditions as described in reference \cite{han2016spectral}. 
In the experiment of reference \cite{han2016spectral}, transmission spectra of the probe light passing through an atomic cloud driven in conditions of Rydberg EIT were acquired as a function of probe detuning $\Delta_p$, while keeping $\Delta_c=0$. A couple of such experimental spectra are shown in Fig. \ref{Figure2Levy} along with spectra calculated using Eq. \eqref{equaIp}.
The spectra simulated from Eq. \eqref{equaIp} with the approximate susceptibility of Eq. \eqref{chiExact} are in good agreement with the experimental ones as shown in Fig. \ref{Figure2Levy}. They are also very similar to the spectra computed from Eq. \eqref{equaIp} with the input of $\chi_{MC}$. 
In conclusion, Eqs. \eqref{equaIp}, \eqref{chiExact}, and \eqref{Rydfrac2} yield reliable and easily computable results, which capture most of the experimental features. It can be verified further that this model predicts the disappearance of the EIT spectral shift for long Gaussian atomic samples, as was observed in Ref.~\cite{han2016spectral}.
There is a slight discrepancy between experimental and simulated data for larger atomic densities, similar to that observed with the enhanced mean-field model of Ref.~\cite{han2016spectral}. It may be due to motion induced dephasing, which should be accounted for with a dynamical model beyond the scope of the frozen gas approximation assumed in this section~\cite{garttner2013nonlinear}.

\section{Summary}

The distribution of energy shifts in a gas of interacting Rydberg atoms was shown to follow a simple L\'evy-type distribution which depends mostly on the Rydberg atom density $n_r$, the interaction strength $C_p$, and the blockade distance $D_b$. Based on this result, we have demonstrated that the recent observation of spectral shift and dephasing of Rydberg EIT spectra in presence of strong Rydberg interactions is well captured by a model where the susceptibility is calculated as a statistical average. 
Spectroscopic measurements of the distribution of energy level shifts in an interacting Rydberg gas should provide a complete test of the theory presented in this article.

\begin{acknowledgments}
We are grateful to Beno\^{\i}t Gr\'{e}maud for helpful discussions. This work is supported by the National Research Foundation, Prime Minister's Office, Singapore and the Ministry of Education, Singapore under the Research Centres of Excellence programme, as well as by Singapore Ministry of Education Academic Research Fund Tier 2 (Grant No. MOE2015-T2-1-085).
\end{acknowledgments}

\appendix
\section{ Analytical derivation of the L\'evy stable distribution \label{Tmodel}}
\subsection{Mathematical definition of L\'evy stable distributions}
A L\'evy stable distribution $g_{\alpha,s}$ with exponent $\alpha \in (0,1)$ and scale $s>0$ is a distribution on $\nu \in (0,\infty)$ with Laplace transform
\begin{align*}
\int_{0}^{\infty} e^{-t \, \nu} \, g_{\alpha,s}(\nu) \, d\nu
\; = \; 
\exp\curBK{-(s \, t)^{\alpha}}.
\end{align*}
For notational convenience, we introduce the parameter $\lambda \equiv s^{\alpha}$, write the Laplace transform as $\cL_{\alpha,\lambda}(t) \equiv \exp\left(-\lambda \, t^{\alpha} \right) $, and define $f_{\alpha,\lambda}(\nu) \equiv g_{\alpha,s}(\nu)$. The generic example is the L\'evy (van der Waals) distribution
\begin{align*}
f_{1/2,\lambda}(\nu) =\frac{\lambda}{2 \sqrt{\pi}}  \nu^{-3/2} \, e^{-\lambda^2/(4 \nu)}.
\end{align*}
Other simple examples are 
$$f_{1/3,\lambda}(\nu)=\frac{\lambda \, \textrm{Ai} \left( \frac{3^{1/3} \lambda}{\nu^{1/3}}\right)}{3^{1/3} \, \nu^{4/3}},$$
where Ai stands for the Airy function, and 
$$f_{2/3,\lambda}(\nu)=\frac{2 \sqrt{3} \lambda^{3}}{27 \pi \nu^3} \textrm{e}^{\frac{-2 \, \lambda^3}{27 \, \nu^2}} \left[ K_{1/3}\left( \frac{2 \, \lambda^3}{27 \, \nu^2}\right) +  K_{2/3}\left( \frac{2 \, \lambda^3}{27 \, \nu^2}\right) \right],$$
 expressed with modified Bessel functions of the second kind \cite{gorska2012levy}.

\subsection{Distribution of energy shifts : point-like particles}

We consider an energy of the type 
\begin{align*}
\textrm{(Energy)} \equiv \sum_{k=1}^n 1/\|X_k\|^p
\end{align*}
where $\{ X_k \}_{k}$ are the positions of $n = n_r \, V$ (independent) atoms uniformly distributed in the ball $B(R) \subset \bR^d$ of radius $R$ and volume $V(R)= d^{-1} \, S_d \,  R^d$; the notation $S_d$ indicates the surface of the unit sphere in $\bR^d$, and we assume that $0<d/p<1$. The Laplace transform of the energy reads (for $t>0$)
\begin{align*}
\cL(t) 
&= 
\BK{V^{-1} \int_{B(R)} e^{-t / r^p} \, d \vecr}^n\\
&=
\BK{1-V^{-1} \int_{B(R)} \BK{1-e^{-t/ r^p}} \, d \vecr}^n\\
&=
\BK{1- \frac{n_r \, S_d}{n}\int_{0}^R \BK{1-e^{-t/ r^p}} \, r^{d-1} \, dr}^n\\
&\to
\exp \curBK{- n_r \, S_d \, \int_{r=0}^\infty \BK{1-e^{-t / r^p}} \, r^{d-1} \, dr}\\
&=
\exp \curBK{- n_r \, S_d\, p^{-1} \, t^{d/p} \, \int_{u=0}^\infty \BK{1-e^{-u}} \, u^{-d/p} \, \frac{du}{u}}\\
&=
\exp \curBK{- \frac{ n_r \, S_d\, \Gamma(1-d/p) }{d} \, t^{d/p} } \equiv \cL_{\frac{d}{p},\lambda_0}(t),
\end{align*}
where the limit was taken for $n \rightarrow \infty$ at fixed density $n_r$.
The final result $\cL_{\frac{d}{p},\lambda_0}(t)$ is the Laplace transform of a L\'evy stable distribution with exponent $\alpha = d/p$ and parameter 
$$\lambda_{0} \equiv \frac{ n_r \, S_d\, \Gamma(1-d/p) }{d}.$$

This result is a particular case of the more general theory of the sum of independent and random variables having `heavytail' probability distributions with power-law dependence \cite{bardou2002levy}.
The probability distribution of the energy is indeed that of a sum of random variables $Y_k=1/\|X_k\|^p$ with power-law dependent probability distributions $$f_{Y_k}(y)=\frac{d }{p \, R^d} \, \frac{1}{y^{1+d/p}}.$$

\subsection{Distribution of energy shifts : hard ball model}

A full analytical derivation of the distribution of energy shifts for non-independent distributions of particles is beyond the scope of this article. We emphasize here the main arguments leading to the derivation of the coefficient $\epsilon$ of Eq. \eqref{AnaResult} in case where the parameter $\eta = n_r \, d^{-1} \, S_d \, D_b^d$ is small and considering a simplified system of hard balls of radii $D_b/2$ in $\bR^d$. The calculation is based on the cluster expansion used for example for the derivation of the first corrections to the equation of state of a van der Waals gas \cite{mayer1941molecular,krauth2006statistical}.

Again, the analytical derivation of the distribution of energy shifts relies on the computation of its Laplace transform as 
\small
\begin{align*}
\cL_\star(t) = 
\frac{1}{Z} \, \int_{B(R)^n} \prod_{j < k} 
\BK{ 1-\gamma(\vecr_k, \vecr_j) }
 \,
e^{ -t \, \sum_{i=1}^n \frac {1}{r_i^p}}  \, d\vecr_1 \ldots d\vecr_n
\end{align*}
\normalsize
where $1-\gamma (\vec r_j , \vec r_k )$ is equal to 0 for $|\vec r_j - \vec r_k| \leq D_b$ and 1 otherwise. $Z$ is the partition function given by :
\small
\begin{equation}
Z=\int_{B(R)^n} \prod_{j<k} \left( 1-\gamma (\vec r_j , \vec r_k )\right) \, d\vecr_1 \ldots d\vecr_n.\notag
\end{equation}
\normalsize
The first order approximation consists in dropping all the products involving two $\gamma$ functions from the integrands, that means 
$$\prod_{j<k} \left( 1-\gamma (\vec r_j , \vec r_k )\right)\approx 1-\sum_{j<k}\gamma (\vec r_j , \vec r_k ).$$
With this approximation, the partition function reduces to 
$$ Z\approx V^n \left(1- \frac{n-1}{2} \eta \right)\equiv V^n \, \BK{1 - \zeta}.$$
 This result is valid in the limit of small enough $\eta$ and $n$ such that the correction to the partition function remains relatively small. Further, the number of atoms  $n$ will be assumed sufficiently large in order to calculate asymptotic behaviors while still having  
 $$\zeta =\frac{n-1}{2} \eta \ll 1.$$ 
This approximation yields that
%\newpage
%
\begin{widetext}
\begin{align*}
\cL_\star(t) 
&\approx 
\frac{1}{V^n \, (1-\zeta)} \, \int_{B(R)^n} \BK{1-\sum_{j < k} 
\gamma(\vecr_k, \vecr_j) } \,
\curBK{ \exp\BK{-t \, \sum_{i=1}^n 1/r_i^p} } \, d\vecr_1 \ldots d\vecr_n \\
&\approx
\frac{1}{1-\zeta} \, \cL_{\frac{d}{p},\lambda_0}(t)
- \frac{n(n-1)}{2(1-\zeta)} 
\curBK{\frac{1}{V^2} \int_{B(R)^2} e^{-t (1 / r_1^p + 1/r_2^p)} \, 1_{|\vecr_1-\vecr_2|<D_b} d \vecr_1 d \vecr_2}  \,  \BK{\frac{1}{V} \int_{B(R)} e^{-t / r^p} \, d \vecr}^{n-2}\\
&\approx
\frac{1}{1-\zeta} \, \cL_{\frac{d}{p},\lambda_0}(t)
\curBK{
1 - 
\frac{n(n-1)}{2V} \, A(t,d,D_b) \, 
\BK{\frac{1}{V} \int_{B(R)} e^{-t / r^p} \, d \vecr}^{-2}
},
\end{align*}
\end{widetext}
where on the one hand we have

\small
\begin{align*}
A(t,d,D_b)
&=
\frac{1}{V} \int_{B(R)^2} e^{-t (1 / r_1^p + 1/r_2^p)} \, 1_{|\vecr_1-\vecr_2|<D_b} d \vecr_1 d \vecr_2\\
&\approx
\frac{S_d \, D_b^d}{d \, V} \int_{B(R)} e^{ -2 \, t / r_1^p } d \vecr_1\\
&=
\frac{S_d \, D_b^d}{d}  \,\curBK{
 1-\frac{n_r \, S_d}{n} \int_{r=0}^\infty \BK{1-e^{-2 \, t / r^p}} \, r^{d-1} \, dr
 }\\
 &=
\frac{S_d \, D_b^d}{d}  \,\curBK{
 1 + \frac{1}{n} \log \cL_{\frac{d}{p},\lambda_0}(2 \, t)
 }\\
 &=
\frac{S_d \, D_b^d}{d}  \,\curBK{
 1 - 2^{d/p} \, \frac{\lambda_{0} \, t^{d/p}}{n}
 },
\end{align*}
\normalsize
and on the other hand and very similarly, we have
\small
\begin{align*}
\BK{ \frac{1}{V} \int_{B(R)} e^{-t / r_1^p } d \vecr_1 }^{-2}
&=
\BK{
 1 + \frac{1}{n} \log \cL_{\frac{d}{p},\lambda_0}(t)
 }^{-2}\\
&\approx
 1 + \frac{2}{n} \, \lambda_{0} \, t^{d/p}.
\end{align*}
\normalsize
Substitution of the last two results back into $\cL_\star(t)$, and using the assumptions $\zeta \ll 1$ and $\zeta/n \approx \eta/2$ yield that
\begin{align*}
\cL_\star(t) 
&\approx
\cL_{\frac{d}{p},\lambda_0}(t) \, \BK{1-\zeta \, \frac{ 2-2^{d/p} }{n} \, \lambda_{0} \, t^{d/p}}\\
&\approx
\cL_{\frac{d}{p},\lambda_0}(t) + \frac{d}{d\lambda} \cL_{\frac{d}{p},\lambda_0}(t) \times \BK{\frac{\zeta}{n} \, (2-2^{d/p}) \, \lambda_{0} }\\
&\approx \cL_{\frac{d}{p},\lambda_{0} \, (1 + \frac{2-2^{d/p}}{2} \, \eta) }(t).
\end{align*}
In other words, in $\bR^d$ and with interaction potentials varying as $1/r^p$, effects of finite ball size on the distribution of energy shifts are entirely accounted for as a shift of the parameter $\lambda_0$,
$$
\lambda_{0} \mapsto 
\lambda_{0} \, \BK{ 1 + \epsilon \, n_r \, V(D_b) }.
$$
with $\epsilon = \frac{2-2^{d/p}}{2}$ and $V(D_b)$ the exclusion volume of a ball of radius $D_b$.

This analytical result seems to hold for somewhat larger $D_b$ than given by the range of validity of the present calculation. Indeed, the condition $n \times \eta/2 \ll 1$ for large $n$ up to 10 is equivalent to $\left(\frac{D_b}{R_d} \right)^d \ll \frac{1}{5}$, i. e. $D_b$ cannot be larger than a small fraction of the interatomic distance $R_d$, whereas the simulation by Markov chain Monte-Carlo of hard balls in Fig. \ref{CorrectionCoefHS} gave a similar correction $\sim \lambda_0 \times \epsilon \, n_r \, V(D_b)$ up to $D_b\approx 0.7\, R_3$.

\smallskip

\section{Comparison between the L\'evy model and Monte Carlo rate equations}
\label{LevyModel}

The aim of this Appendix is to come back to the different approaches for computing the imaginary part of the susceptibility as in Fig. \ref{AppendixFig}.
MCRE is based on the computation of single atom transition rates that depend on the Rydberg energy shift of each atom in the presence of surrounding atoms already excited to Rydberg state. 
With the ladder configuration of excitation with near-resonant lasers of Fig. \ref{SetupRydEIT}, performing the adiabatic elimination of atomic coherences on the master equation in order to derive rate equations suitable for Monte Carlo simulations fails as it may yield negative transition rates even for a non interacting system.
For simulations that are interested only in steady state solutions, this problem can be overcome by considering strictly positive transitions rates that evolve a system of non-interacting atoms into the correct single atom steady states but does not reproduce properly the states of the system at intermediate time steps \cite{ates2011electromagnetically,GarrtnerSemianalytical2014}. More precisely, the state vector of the single atom populations, $\rho(t)=\left(\rho_{gg},\rho_{ee},\rho_{rr} \right)^T$ is assumed to obey rate equations 
$$\frac{d \rho}{dt}=B \rho,$$
 where $B$ is a well-designed matrix having positive off diagonal elements ensuring transition rates between two different states are always positive, and $\rho(\infty)=\left(\rho_{gg}^{(S)},\rho_{ee}^{(S)},\rho_{rr}^{(S)} \right)^T$ equals the steady state populations of the single atom master equation. This rate equations solution is readily extended to the case of an ensemble of interacting atoms by calculating the effect of energy shifts on the matrix $B$, locally and at each time step of the computation, i. e. computing the effect of the detuning change $-2 \pi \nu_{i}$ of laser (2) in Fig. \ref{SetupRydEIT}, where $\nu_i$ is given in Eq.~\eqref{freqchangeatomic}.

 In Fig. \ref{AppendixFig}, the intermediate state population fraction $f_e$ is used in order to calculate the susceptibility from the simple formula 
 $$\mathrm{Im} \left[ \chi_{MC}\right]=f_e\,  \frac{\hbar c n_{at}} {I_p \Gamma_e},$$
  and is obtained directly from MCRE computation with the $B$ matrix of reference \cite{heeg2012hybrid}
 \[B=
  \begin{bmatrix}
    \rho_{gg}^{(S)}-1 & \rho_{gg}^{(S)} & \rho_{gg}^{(S)} \\
    \rho_{ee}^{(S)} & \rho_{ee}^{(S)}-1 & \rho_{ee}^{(S)} \\
    \rho_{rr}^{(S)} & \rho_{rr}^{(S)} & \rho_{rr}^{(S)}-1
  \end{bmatrix}
\] 
This simple form of the matrix tends to define very large transition rates. Other matrices exist, exhibiting transition rates yielding time evolutions of the single-atom populations that are closer to the real ones \cite{ates2011electromagnetically}.

In Fig.  \ref{AppendixFig} is also plotted $f_e=\frac{I_p \Gamma_e} {\hbar c n_{at}}\mathrm{Im} \left[ \chi\right]$, where $\mathrm{Im} \left[ \chi\right]$ writes
\begin{equation}
\mathrm{Im}\left[\chi\right]=\left( 1- f_r \right) \int_0^{+\infty} \mathrm{Im}\left[\chi_0 (\nu) \right] f (\nu) d\nu,
\label{chiExact2}
\end{equation}
In this expression of $\mathrm{Im}\left[\chi\right]$, $f(\nu)$ is a weighting distribution depending on $\nu$, the local shift of the Rydberg energy level, $f_r$ is the Rydberg atom fraction, while $\chi_0$ is the non interacting 3-level atom susceptibility for the probe light. $\chi_0$ is given as below
 
\begin{align}
\chi_0 = -\frac{2 n_{at} d_{eg}^2}{\varepsilon_0 \hbar \Omega_p} \vroW_{eg},
\label{chi}
\end{align}
where $d_{eg}$ is the dipole matrix element for the $|g\rangle\rightarrow|e\rangle$ transition, $n_{at}$ is the atomic density, assumed locally  uniform, and $\Omega_{p}  = -d_{eg} \mc{E}_{p} / \hbar$ is the Rabi frequency of the probe field. $\vroW_{eg}$ is the atomic coherence, given by the solution of the single-body master equation
$$\partial_t \vroW = - \frac{i}{\hbar} [ h , \vroW ] +\mc{L} \vroW,$$
where $h$ is the Hamiltonian of interaction with the probe and coupling lights. $\mc{L}\vroW$ is a set of Liouvillians simulating the effects of the dissipative processes
\begin{align}
\mc{L}\vroW \approx & -  \frac{\Gamma_e}{2} \left( \ket{e}\bra{e}  \vroW  + \vroW \ket{e}\bra{e}- 2 \ket{g}\bra{e} \vroW \ket{e}\bra{g} \right) \notag  \\
& -  \gamma_{d} \left( \ket{g}\bra{g}  \vroW \ket{r}\bra{r}  + \ket{r}\bra{r} \vroW \ket{g}\bra{g}  \right) \notag\\
& -  \gamma_{d} \left( \ket{e}\bra{e}  \vroW \ket{r}\bra{r}  + \ket{r}\bra{r} \vroW \ket{e}\bra{e}  \right). \notag
\end{align}
The atomic coherences $\varrho_{gr}$ and $\varrho_{rg}$ decay with a rate $\gamma_{d}$ which is of the order of $\gamma_0 / 2 \pi=0.1$~MHz in absence of interactions.
 In Fig. \ref{Figure2Levy}, we use a slightly larger value for the decay rate of atomic coherence, $\gamma_{d} / 2 \pi=0.6$~MHz, in order to account for residual effects of motional dephasing induced by dipolar forces. The Rydberg atom fraction $f_r$ in Eq. \eqref{chiExact2} may be obtained via Eq. \eqref{Rydfrac2} from the knowledge of $f_0  \equiv \vroW_{rr}\left( \nu=0 \right)$, the Rydberg atom fraction absent any interaction, and $V_B$ the volume inside a blockade sphere.
In our simulations, Eqs. \eqref{chiExact2} and \eqref{Rydfrac2} are solved with $\chi_0$, $f_0$, and $V_B$ calculated to all orders in $\Omega_p$.

\begin{thebibliography}{39}%
\makeatletter
\providecommand \@ifxundefined [1]{%
 \@ifx{#1\undefined}
}%
\providecommand \@ifnum [1]{%
 \ifnum #1\expandafter \@firstoftwo
 \else \expandafter \@secondoftwo
 \fi
}%
\providecommand \@ifx [1]{%
 \ifx #1\expandafter \@firstoftwo
 \else \expandafter \@secondoftwo
 \fi
}%
\providecommand \natexlab [1]{#1}%
\providecommand \enquote  [1]{``#1''}%
\providecommand \bibnamefont  [1]{#1}%
\providecommand \bibfnamefont [1]{#1}%
\providecommand \citenamefont [1]{#1}%
\providecommand \href@noop [0]{\@secondoftwo}%
\providecommand \href [0]{\begingroup \@sanitize@url \@href}%
\providecommand \@href[1]{\@@startlink{#1}\@@href}%
\providecommand \@@href[1]{\endgroup#1\@@endlink}%
\providecommand \@sanitize@url [0]{\catcode `\\12\catcode `\$12\catcode
  `\&12\catcode `\#12\catcode `\^12\catcode `\_12\catcode `\%12\relax}%
\providecommand \@@startlink[1]{}%
\providecommand \@@endlink[0]{}%
\providecommand \url  [0]{\begingroup\@sanitize@url \@url }%
\providecommand \@url [1]{\endgroup\@href {#1}{\urlprefix }}%
\providecommand \urlprefix  [0]{URL }%
\providecommand \Eprint [0]{\href }%
\providecommand \doibase [0]{http://dx.doi.org/}%
\providecommand \selectlanguage [0]{\@gobble}%
\providecommand \bibinfo  [0]{\@secondoftwo}%
\providecommand \bibfield  [0]{\@secondoftwo}%
\providecommand \translation [1]{[#1]}%
\providecommand \BibitemOpen [0]{}%
\providecommand \bibitemStop [0]{}%
\providecommand \bibitemNoStop [0]{.\EOS\space}%
\providecommand \EOS [0]{\spacefactor3000\relax}%
\providecommand \BibitemShut  [1]{\csname bibitem#1\endcsname}%
\let\auto@bib@innerbib\@empty
%</preamble>
\bibitem [{\citenamefont {Bardou}\ \emph {et~al.}(2002)\citenamefont {Bardou},
  \citenamefont {Bouchaud}, \citenamefont {Aspect},\ and\ \citenamefont
  {Cohen-Tannoudji}}]{bardou2002levy}%
  \BibitemOpen
  \bibfield  {author} {\bibinfo {author} {\bibfnamefont {F.}~\bibnamefont
  {Bardou}}, \bibinfo {author} {\bibfnamefont {J.-P.}\ \bibnamefont
  {Bouchaud}}, \bibinfo {author} {\bibfnamefont {A.}~\bibnamefont {Aspect}}, \
  and\ \bibinfo {author} {\bibfnamefont {C.}~\bibnamefont {Cohen-Tannoudji}},\
  }\href@noop {} {\emph {\bibinfo {title} {L{\'e}vy statistics and laser
  cooling: how rare events bring atoms to rest}}}\ (\bibinfo  {publisher}
  {Cambridge University Press},\ \bibinfo {year} {2002})\BibitemShut {NoStop}%
\bibitem [{\citenamefont {Ott}\ \emph {et~al.}(1990)\citenamefont {Ott},
  \citenamefont {Bouchaud}, \citenamefont {Langevin},\ and\ \citenamefont
  {Urbach}}]{ott1990anomalous}%
  \BibitemOpen
  \bibfield  {author} {\bibinfo {author} {\bibfnamefont {A.}~\bibnamefont
  {Ott}}, \bibinfo {author} {\bibfnamefont {J.~P.}\ \bibnamefont {Bouchaud}},
  \bibinfo {author} {\bibfnamefont {D.}~\bibnamefont {Langevin}}, \ and\
  \bibinfo {author} {\bibfnamefont {W.}~\bibnamefont {Urbach}},\ }\bibfield
  {title} {\enquote {\bibinfo {title} {Anomalous diffusion in living polymers:
  A genuine levy flight?}}\ }\href@noop {} {\bibfield  {journal} {\bibinfo
  {journal} {Phys. Rev. Lett.}\ }\textbf {\bibinfo {volume} {65}},\ \bibinfo
  {pages} {2201} (\bibinfo {year} {1990})}\BibitemShut {NoStop}%
\bibitem [{\citenamefont {Edwards}\ \emph {et~al.}(2007)\citenamefont
  {Edwards}, \citenamefont {Phillips}, \citenamefont {Watkins}, \citenamefont
  {Freeman}, \citenamefont {Murphy}, \citenamefont {Afanasyev}, \citenamefont
  {Buldyrev}, \citenamefont {da~Luz}, \citenamefont {Raposo}, \citenamefont
  {Stanley} \emph {et~al.}}]{edwards2007revisiting}%
  \BibitemOpen
  \bibfield  {author} {\bibinfo {author} {\bibfnamefont {A.~M.}\ \bibnamefont
  {Edwards}}, \bibinfo {author} {\bibfnamefont {R.~A.}\ \bibnamefont
  {Phillips}}, \bibinfo {author} {\bibfnamefont {N.~W.}\ \bibnamefont
  {Watkins}}, \bibinfo {author} {\bibfnamefont {M.~P.}\ \bibnamefont
  {Freeman}}, \bibinfo {author} {\bibfnamefont {E.~J.}\ \bibnamefont {Murphy}},
  \bibinfo {author} {\bibfnamefont {V.}~\bibnamefont {Afanasyev}}, \bibinfo
  {author} {\bibfnamefont {S.~V.}\ \bibnamefont {Buldyrev}}, \bibinfo {author}
  {\bibfnamefont {M.~G.~E.}\ \bibnamefont {da~Luz}}, \bibinfo {author}
  {\bibfnamefont {E.~P.}\ \bibnamefont {Raposo}}, \bibinfo {author}
  {\bibfnamefont {H.~E.}\ \bibnamefont {Stanley}},  \emph {et~al.},\ }\bibfield
   {title} {\enquote {\bibinfo {title} {Revisiting l{\'e}vy flight search
  patterns of wandering albatrosses, bumblebees and deer},}\ }\href@noop {}
  {\bibfield  {journal} {\bibinfo  {journal} {Nature}\ }\textbf {\bibinfo
  {volume} {449}},\ \bibinfo {pages} {1044--1048} (\bibinfo {year}
  {2007})}\BibitemShut {NoStop}%
\bibitem [{\citenamefont {Rachev}\ \emph {et~al.}(2011)\citenamefont {Rachev},
  \citenamefont {Kim}, \citenamefont {Bianchi},\ and\ \citenamefont
  {Fabozzi}}]{rachev2011financial}%
  \BibitemOpen
  \bibfield  {author} {\bibinfo {author} {\bibfnamefont {S.~T.}\ \bibnamefont
  {Rachev}}, \bibinfo {author} {\bibfnamefont {Y.~S.}\ \bibnamefont {Kim}},
  \bibinfo {author} {\bibfnamefont {M.~L.}\ \bibnamefont {Bianchi}}, \ and\
  \bibinfo {author} {\bibfnamefont {F.~J.}\ \bibnamefont {Fabozzi}},\
  }\href@noop {} {\emph {\bibinfo {title} {Financial models with L{\'e}vy
  processes and volatility clustering}}},\ Vol.\ \bibinfo {volume} {187}\
  (\bibinfo  {publisher} {John Wiley \& Sons},\ \bibinfo {year}
  {2011})\BibitemShut {NoStop}%
\bibitem [{\citenamefont {Ibragimov}\ \emph {et~al.}(2015)\citenamefont
  {Ibragimov}, \citenamefont {Ibragimov},\ and\ \citenamefont
  {Walden}}]{ibragimov2015heavy}%
  \BibitemOpen
  \bibfield  {author} {\bibinfo {author} {\bibfnamefont {M.}~\bibnamefont
  {Ibragimov}}, \bibinfo {author} {\bibfnamefont {R.}~\bibnamefont
  {Ibragimov}}, \ and\ \bibinfo {author} {\bibfnamefont {J.}~\bibnamefont
  {Walden}},\ }\href@noop {} {\emph {\bibinfo {title} {Heavy-tailed
  distributions and robustness in economics and finance}}},\ Vol.\ \bibinfo
  {volume} {214}\ (\bibinfo  {publisher} {Springer},\ \bibinfo {year}
  {2015})\BibitemShut {NoStop}%
\bibitem [{\citenamefont {Shlesinger}\ and\ \citenamefont
  {Frisch~(eds)}(1995)}]{Shlesinger1995levy}%
  \BibitemOpen
  \bibfield  {author} {\bibinfo {author} {\bibfnamefont {Zaslavsky G.~M.}\
  \bibnamefont {Shlesinger}, \bibfnamefont {M.~F.}}\ and\ \bibinfo {author}
  {\bibfnamefont {U.}~\bibnamefont {Frisch~(eds)}},\ }\href@noop {} {\emph
  {\bibinfo {title} {L{\'e}vy flights and related topics in physics}}}\
  (\bibinfo  {publisher} {Springer},\ \bibinfo {year} {1995})\BibitemShut
  {NoStop}%
\bibitem [{\citenamefont {Barthelemy}\ \emph {et~al.}(2008)\citenamefont
  {Barthelemy}, \citenamefont {Bertolotti},\ and\ \citenamefont
  {Wiersma}}]{barthelemy2008levy}%
  \BibitemOpen
  \bibfield  {author} {\bibinfo {author} {\bibfnamefont {P.}~\bibnamefont
  {Barthelemy}}, \bibinfo {author} {\bibfnamefont {J.}~\bibnamefont
  {Bertolotti}}, \ and\ \bibinfo {author} {\bibfnamefont {D.~S.}\ \bibnamefont
  {Wiersma}},\ }\bibfield  {title} {\enquote {\bibinfo {title} {A l{\'e}vy
  flight for light},}\ }\href@noop {} {\bibfield  {journal} {\bibinfo
  {journal} {Nature}\ }\textbf {\bibinfo {volume} {453}},\ \bibinfo {pages}
  {495--498} (\bibinfo {year} {2008})}\BibitemShut {NoStop}%
\bibitem [{\citenamefont {Zaburdaev}\ \emph {et~al.}(2015)\citenamefont
  {Zaburdaev}, \citenamefont {Denisov},\ and\ \citenamefont
  {Klafter}}]{ZaburdaevLevy2015}%
  \BibitemOpen
  \bibfield  {author} {\bibinfo {author} {\bibfnamefont {V.}~\bibnamefont
  {Zaburdaev}}, \bibinfo {author} {\bibfnamefont {S.}~\bibnamefont {Denisov}},
  \ and\ \bibinfo {author} {\bibfnamefont {J.}~\bibnamefont {Klafter}},\
  }\bibfield  {title} {\enquote {\bibinfo {title} {L\'evy walks},}\ }\href
  {\doibase 10.1103/RevModPhys.87.483} {\bibfield  {journal} {\bibinfo
  {journal} {Rev. Mod. Phys.}\ }\textbf {\bibinfo {volume} {87}},\ \bibinfo
  {pages} {483--530} (\bibinfo {year} {2015})}\BibitemShut {NoStop}%
\bibitem [{\citenamefont {Margenau}(1935)}]{margenau1935theory}%
  \BibitemOpen
  \bibfield  {author} {\bibinfo {author} {\bibfnamefont {H.}~\bibnamefont
  {Margenau}},\ }\bibfield  {title} {\enquote {\bibinfo {title} {Theory of
  pressure effects of foreign gases on spectral lines},}\ }\href@noop {}
  {\bibfield  {journal} {\bibinfo  {journal} {Physical Review}\ }\textbf
  {\bibinfo {volume} {48}},\ \bibinfo {pages} {755} (\bibinfo {year}
  {1935})}\BibitemShut {NoStop}%
\bibitem [{\citenamefont {Ates}\ \emph {et~al.}(2007)\citenamefont {Ates},
  \citenamefont {Pohl}, \citenamefont {Pattard},\ and\ \citenamefont
  {Rost}}]{ates2007antiblockade}%
  \BibitemOpen
  \bibfield  {author} {\bibinfo {author} {\bibfnamefont {C.}~\bibnamefont
  {Ates}}, \bibinfo {author} {\bibfnamefont {T.}~\bibnamefont {Pohl}}, \bibinfo
  {author} {\bibfnamefont {T.}~\bibnamefont {Pattard}}, \ and\ \bibinfo
  {author} {\bibfnamefont {J.~M.}\ \bibnamefont {Rost}},\ }\bibfield  {title}
  {\enquote {\bibinfo {title} {Antiblockade in rydberg excitation of an
  ultracold lattice gas},}\ }\href@noop {} {\bibfield  {journal} {\bibinfo
  {journal} {Phys. Rev. Lett.}\ }\textbf {\bibinfo {volume} {98}},\ \bibinfo
  {pages} {023002} (\bibinfo {year} {2007})}\BibitemShut {NoStop}%
\bibitem [{\citenamefont {Vogt}\ \emph {et~al.}(2007)\citenamefont {Vogt},
  \citenamefont {Viteau}, \citenamefont {Chotia}, \citenamefont {Zhao},
  \citenamefont {Comparat},\ and\ \citenamefont {Pillet}}]{vogt2007electric}%
  \BibitemOpen
  \bibfield  {author} {\bibinfo {author} {\bibfnamefont {T.}~\bibnamefont
  {Vogt}}, \bibinfo {author} {\bibfnamefont {M.}~\bibnamefont {Viteau}},
  \bibinfo {author} {\bibfnamefont {A.}~\bibnamefont {Chotia}}, \bibinfo
  {author} {\bibfnamefont {J.}~\bibnamefont {Zhao}}, \bibinfo {author}
  {\bibfnamefont {D.}~\bibnamefont {Comparat}}, \ and\ \bibinfo {author}
  {\bibfnamefont {P.}~\bibnamefont {Pillet}},\ }\bibfield  {title} {\enquote
  {\bibinfo {title} {Electric-field induced dipole blockade with rydberg
  atoms},}\ }\href@noop {} {\bibfield  {journal} {\bibinfo  {journal} {Phys.
  Rev. Lett.}\ }\textbf {\bibinfo {volume} {99}},\ \bibinfo {pages} {073002}
  (\bibinfo {year} {2007})}\BibitemShut {NoStop}%
\bibitem [{\citenamefont {Chotia}\ \emph {et~al.}(2008)\citenamefont {Chotia},
  \citenamefont {Viteau}, \citenamefont {Vogt}, \citenamefont {Comparat},\ and\
  \citenamefont {Pillet}}]{chotia2008kinetic}%
  \BibitemOpen
  \bibfield  {author} {\bibinfo {author} {\bibfnamefont {A.}~\bibnamefont
  {Chotia}}, \bibinfo {author} {\bibfnamefont {M.}~\bibnamefont {Viteau}},
  \bibinfo {author} {\bibfnamefont {T.}~\bibnamefont {Vogt}}, \bibinfo {author}
  {\bibfnamefont {D.}~\bibnamefont {Comparat}}, \ and\ \bibinfo {author}
  {\bibfnamefont {P.}~\bibnamefont {Pillet}},\ }\bibfield  {title} {\enquote
  {\bibinfo {title} {Kinetic monte carlo modeling of dipole blockade in rydberg
  excitation experiment},}\ }\href@noop {} {\bibfield  {journal} {\bibinfo
  {journal} {New. J. Phys.}\ }\textbf {\bibinfo {volume} {10}},\ \bibinfo
  {pages} {045031} (\bibinfo {year} {2008})}\BibitemShut {NoStop}%
\bibitem [{\citenamefont {Ates}\ \emph {et~al.}(2011)\citenamefont {Ates},
  \citenamefont {Sevin{\c{c}}li},\ and\ \citenamefont
  {Pohl}}]{ates2011electromagnetically}%
  \BibitemOpen
  \bibfield  {author} {\bibinfo {author} {\bibfnamefont {C.}~\bibnamefont
  {Ates}}, \bibinfo {author} {\bibfnamefont {S.}~\bibnamefont
  {Sevin{\c{c}}li}}, \ and\ \bibinfo {author} {\bibfnamefont {T.}~\bibnamefont
  {Pohl}},\ }\bibfield  {title} {\enquote {\bibinfo {title}
  {Electromagnetically induced transparency in strongly interacting rydberg
  gases},}\ }\href@noop {} {\bibfield  {journal} {\bibinfo  {journal} {Phys.
  Rev. A}\ }\textbf {\bibinfo {volume} {83}},\ \bibinfo {pages} {041802}
  (\bibinfo {year} {2011})}\BibitemShut {NoStop}%
\bibitem [{\citenamefont {Heeg}\ \emph {et~al.}(2012)\citenamefont {Heeg},
  \citenamefont {G{\"a}rttner},\ and\ \citenamefont {Evers}}]{heeg2012hybrid}%
  \BibitemOpen
  \bibfield  {author} {\bibinfo {author} {\bibfnamefont {K.~P.}\ \bibnamefont
  {Heeg}}, \bibinfo {author} {\bibfnamefont {M.}~\bibnamefont {G{\"a}rttner}},
  \ and\ \bibinfo {author} {\bibfnamefont {J.}~\bibnamefont {Evers}},\
  }\bibfield  {title} {\enquote {\bibinfo {title} {Hybrid model for rydberg
  gases including exact two-body correlations},}\ }\href@noop {} {\bibfield
  {journal} {\bibinfo  {journal} {Phys. Rev. A}\ }\textbf {\bibinfo {volume}
  {86}},\ \bibinfo {pages} {063421} (\bibinfo {year} {2012})}\BibitemShut
  {NoStop}%
\bibitem [{\citenamefont {G{\"a}rttner}\ \emph {et~al.}(2012)\citenamefont
  {G{\"a}rttner}, \citenamefont {Heeg}, \citenamefont {Gasenzer},\ and\
  \citenamefont {Evers}}]{garttner2012finite}%
  \BibitemOpen
  \bibfield  {author} {\bibinfo {author} {\bibfnamefont {M.}~\bibnamefont
  {G{\"a}rttner}}, \bibinfo {author} {\bibfnamefont {K.~P}\ \bibnamefont
  {Heeg}}, \bibinfo {author} {\bibfnamefont {T.}~\bibnamefont {Gasenzer}}, \
  and\ \bibinfo {author} {\bibfnamefont {J.}~\bibnamefont {Evers}},\ }\bibfield
   {title} {\enquote {\bibinfo {title} {Finite-size effects in strongly
  interacting rydberg gases},}\ }\href@noop {} {\bibfield  {journal} {\bibinfo
  {journal} {Phys. Rev. A}\ }\textbf {\bibinfo {volume} {86}},\ \bibinfo
  {pages} {033422} (\bibinfo {year} {2012})}\BibitemShut {NoStop}%
\bibitem [{\citenamefont {Petrosyan}\ \emph {et~al.}(2013)\citenamefont
  {Petrosyan}, \citenamefont {H\"oning},\ and\ \citenamefont
  {Fleischhauer}}]{PetrosyanCorrelations2013}%
  \BibitemOpen
  \bibfield  {author} {\bibinfo {author} {\bibfnamefont {D.}~\bibnamefont
  {Petrosyan}}, \bibinfo {author} {\bibfnamefont {M.}~\bibnamefont {H\"oning}},
  \ and\ \bibinfo {author} {\bibfnamefont {M.}~\bibnamefont {Fleischhauer}},\
  }\bibfield  {title} {\enquote {\bibinfo {title} {Spatial correlations of
  rydberg excitations in optically driven atomic ensembles},}\ }\href {\doibase
  10.1103/PhysRevA.87.053414} {\bibfield  {journal} {\bibinfo  {journal} {Phys.
  Rev. A}\ }\textbf {\bibinfo {volume} {87}},\ \bibinfo {pages} {053414}
  (\bibinfo {year} {2013})}\BibitemShut {NoStop}%
\bibitem [{\citenamefont {G\"arttner}\ \emph {et~al.}(2014)\citenamefont
  {G\"arttner}, \citenamefont {Whitlock}, \citenamefont {Sch\"onleber},\ and\
  \citenamefont {Evers}}]{GarrtnerSemianalytical2014}%
  \BibitemOpen
  \bibfield  {author} {\bibinfo {author} {\bibfnamefont {M.}~\bibnamefont
  {G\"arttner}}, \bibinfo {author} {\bibfnamefont {S.}~\bibnamefont
  {Whitlock}}, \bibinfo {author} {\bibfnamefont {D.~W.}\ \bibnamefont
  {Sch\"onleber}}, \ and\ \bibinfo {author} {\bibfnamefont {J.}~\bibnamefont
  {Evers}},\ }\bibfield  {title} {\enquote {\bibinfo {title} {Semianalytical
  model for nonlinear absorption in strongly interacting rydberg gases},}\
  }\href {\doibase 10.1103/PhysRevA.89.063407} {\bibfield  {journal} {\bibinfo
  {journal} {Phys. Rev. A}\ }\textbf {\bibinfo {volume} {89}},\ \bibinfo
  {pages} {063407} (\bibinfo {year} {2014})}\BibitemShut {NoStop}%
\bibitem [{\citenamefont {Jaksch}\ \emph {et~al.}(2000)\citenamefont {Jaksch},
  \citenamefont {Cirac}, \citenamefont {Zoller}, \citenamefont {Rolston},
  \citenamefont {Cote},\ and\ \citenamefont {Lukin}}]{jaksch:00}%
  \BibitemOpen
  \bibfield  {author} {\bibinfo {author} {\bibfnamefont {D.}~\bibnamefont
  {Jaksch}}, \bibinfo {author} {\bibfnamefont {J.~I.}\ \bibnamefont {Cirac}},
  \bibinfo {author} {\bibfnamefont {P.}~\bibnamefont {Zoller}}, \bibinfo
  {author} {\bibfnamefont {S.~L.}\ \bibnamefont {Rolston}}, \bibinfo {author}
  {\bibfnamefont {R.}~\bibnamefont {Cote}}, \ and\ \bibinfo {author}
  {\bibfnamefont {M.~D.}\ \bibnamefont {Lukin}},\ }\bibfield  {title} {\enquote
  {\bibinfo {title} {Dipole blockade and quantum information processing in
  mesoscopic atomic ensembles},}\ }\href@noop {} {\bibfield  {journal}
  {\bibinfo  {journal} {Phys. Rev. Lett.}\ }\textbf {\bibinfo {volume} {85}},\
  \bibinfo {pages} {2208} (\bibinfo {year} {2000})}\BibitemShut {NoStop}%
\bibitem [{\citenamefont {Tong}\ \emph {et~al.}(2004)\citenamefont {Tong},
  \citenamefont {Farooqi}, \citenamefont {Stanojevic}, \citenamefont
  {Krishnan}, \citenamefont {Zhang}, \citenamefont {C{\^o}t{\'e}},
  \citenamefont {Eyler},\ and\ \citenamefont {Gould}}]{tong2004local}%
  \BibitemOpen
  \bibfield  {author} {\bibinfo {author} {\bibfnamefont {D.}~\bibnamefont
  {Tong}}, \bibinfo {author} {\bibfnamefont {S.~M.}\ \bibnamefont {Farooqi}},
  \bibinfo {author} {\bibfnamefont {J.}~\bibnamefont {Stanojevic}}, \bibinfo
  {author} {\bibfnamefont {S.}~\bibnamefont {Krishnan}}, \bibinfo {author}
  {\bibfnamefont {Y.~P.}\ \bibnamefont {Zhang}}, \bibinfo {author}
  {\bibfnamefont {R.}~\bibnamefont {C{\^o}t{\'e}}}, \bibinfo {author}
  {\bibfnamefont {E.~E.}\ \bibnamefont {Eyler}}, \ and\ \bibinfo {author}
  {\bibfnamefont {P.~L.}\ \bibnamefont {Gould}},\ }\bibfield  {title} {\enquote
  {\bibinfo {title} {Local blockade of rydberg excitation in an ultracold
  gas},}\ }\href@noop {} {\bibfield  {journal} {\bibinfo  {journal} {Phys. Rev.
  Lett.}\ }\textbf {\bibinfo {volume} {93}},\ \bibinfo {pages} {063001}
  (\bibinfo {year} {2004})}\BibitemShut {NoStop}%
\bibitem [{\citenamefont {Vogt}\ \emph {et~al.}(2006)\citenamefont {Vogt},
  \citenamefont {Viteau}, \citenamefont {Zhao}, \citenamefont {Chotia},
  \citenamefont {Comparat},\ and\ \citenamefont {Pillet}}]{vogt2006dipole}%
  \BibitemOpen
  \bibfield  {author} {\bibinfo {author} {\bibfnamefont {T.}~\bibnamefont
  {Vogt}}, \bibinfo {author} {\bibfnamefont {M.}~\bibnamefont {Viteau}},
  \bibinfo {author} {\bibfnamefont {J.}~\bibnamefont {Zhao}}, \bibinfo {author}
  {\bibfnamefont {A.}~\bibnamefont {Chotia}}, \bibinfo {author} {\bibfnamefont
  {D.}~\bibnamefont {Comparat}}, \ and\ \bibinfo {author} {\bibfnamefont
  {P.}~\bibnamefont {Pillet}},\ }\bibfield  {title} {\enquote {\bibinfo {title}
  {Dipole blockade at f{\"o}rster resonances in high resolution laser
  excitation of rydberg states of cesium atoms},}\ }\href@noop {} {\bibfield
  {journal} {\bibinfo  {journal} {Phys. Rev. Lett.}\ }\textbf {\bibinfo
  {volume} {97}},\ \bibinfo {pages} {083003} (\bibinfo {year}
  {2006})}\BibitemShut {NoStop}%
\bibitem [{\citenamefont {Heidemann}\ \emph {et~al.}(2007)\citenamefont
  {Heidemann}, \citenamefont {Raitzsch}, \citenamefont {Bendkowsky},
  \citenamefont {Butscher}, \citenamefont {L\"{o}w}, \citenamefont {Santos},\
  and\ \citenamefont {Pfau}}]{heidemann2007strongblockade}%
  \BibitemOpen
  \bibfield  {author} {\bibinfo {author} {\bibfnamefont {R.}~\bibnamefont
  {Heidemann}}, \bibinfo {author} {\bibfnamefont {U.}~\bibnamefont {Raitzsch}},
  \bibinfo {author} {\bibfnamefont {V.}~\bibnamefont {Bendkowsky}}, \bibinfo
  {author} {\bibfnamefont {B.}~\bibnamefont {Butscher}}, \bibinfo {author}
  {\bibfnamefont {R.}~\bibnamefont {L\"{o}w}}, \bibinfo {author} {\bibfnamefont
  {L.}~\bibnamefont {Santos}}, \ and\ \bibinfo {author} {\bibfnamefont
  {T.}~\bibnamefont {Pfau}},\ }\bibfield  {title} {\enquote {\bibinfo {title}
  {Evidence for coherent collective rydberg excitation in the strong blockade
  regime},}\ }\href@noop {} {\bibfield  {journal} {\bibinfo  {journal} {Phys.
  Rev. Lett.}\ }\textbf {\bibinfo {volume} {99}},\ \bibinfo {pages} {163601}
  (\bibinfo {year} {2007})}\BibitemShut {NoStop}%
\bibitem [{\citenamefont {Pohl}\ \emph {et~al.}(2010)\citenamefont {Pohl},
  \citenamefont {Demler},\ and\ \citenamefont {Lukin}}]{pohl2010dynamical}%
  \BibitemOpen
  \bibfield  {author} {\bibinfo {author} {\bibfnamefont {T.}~\bibnamefont
  {Pohl}}, \bibinfo {author} {\bibfnamefont {E.}~\bibnamefont {Demler}}, \ and\
  \bibinfo {author} {\bibfnamefont {M.~D.}\ \bibnamefont {Lukin}},\ }\bibfield
  {title} {\enquote {\bibinfo {title} {Dynamical crystallization in the dipole
  blockade of ultracold atoms},}\ }\href@noop {} {\bibfield  {journal}
  {\bibinfo  {journal} {Phys. Rev. Lett.}\ }\textbf {\bibinfo {volume} {104}},\
  \bibinfo {pages} {043002} (\bibinfo {year} {2010})}\BibitemShut {NoStop}%
\bibitem [{\citenamefont {Schwarzkopf}\ \emph {et~al.}(2011)\citenamefont
  {Schwarzkopf}, \citenamefont {Sapiro},\ and\ \citenamefont
  {Raithel}}]{schwarzkopf2011imaging}%
  \BibitemOpen
  \bibfield  {author} {\bibinfo {author} {\bibfnamefont {A.}~\bibnamefont
  {Schwarzkopf}}, \bibinfo {author} {\bibfnamefont {R.~E.}\ \bibnamefont
  {Sapiro}}, \ and\ \bibinfo {author} {\bibfnamefont {G.}~\bibnamefont
  {Raithel}},\ }\bibfield  {title} {\enquote {\bibinfo {title} {Imaging spatial
  correlations of rydberg excitations in cold atom clouds},}\ }\href@noop {}
  {\bibfield  {journal} {\bibinfo  {journal} {Phys. Rev. Lett.}\ }\textbf
  {\bibinfo {volume} {107}},\ \bibinfo {pages} {103001} (\bibinfo {year}
  {2011})}\BibitemShut {NoStop}%
\bibitem [{\citenamefont {Schau{\ss}}\ \emph {et~al.}(2012)\citenamefont
  {Schau{\ss}}, \citenamefont {Cheneau}, \citenamefont {Endres}, \citenamefont
  {Fukuhara}, \citenamefont {Hild}, \citenamefont {Omran}, \citenamefont
  {Pohl}, \citenamefont {Gross}, \citenamefont {Kuhr},\ and\ \citenamefont
  {Bloch}}]{schauss2012observation}%
  \BibitemOpen
  \bibfield  {author} {\bibinfo {author} {\bibfnamefont {P.}~\bibnamefont
  {Schau{\ss}}}, \bibinfo {author} {\bibfnamefont {M.}~\bibnamefont {Cheneau}},
  \bibinfo {author} {\bibfnamefont {M.}~\bibnamefont {Endres}}, \bibinfo
  {author} {\bibfnamefont {T.}~\bibnamefont {Fukuhara}}, \bibinfo {author}
  {\bibfnamefont {S.}~\bibnamefont {Hild}}, \bibinfo {author} {\bibfnamefont
  {A.}~\bibnamefont {Omran}}, \bibinfo {author} {\bibfnamefont
  {T.}~\bibnamefont {Pohl}}, \bibinfo {author} {\bibfnamefont {C.}~\bibnamefont
  {Gross}}, \bibinfo {author} {\bibfnamefont {S.}~\bibnamefont {Kuhr}}, \ and\
  \bibinfo {author} {\bibfnamefont {I.}~\bibnamefont {Bloch}},\ }\bibfield
  {title} {\enquote {\bibinfo {title} {Observation of spatially ordered
  structures in a two-dimensional rydberg gas},}\ }\href@noop {} {\bibfield
  {journal} {\bibinfo  {journal} {Nature}\ }\textbf {\bibinfo {volume} {491}},\
  \bibinfo {pages} {87--91} (\bibinfo {year} {2012})}\BibitemShut {NoStop}%
\bibitem [{\citenamefont {Han}\ \emph {et~al.}(2016)\citenamefont {Han},
  \citenamefont {Vogt},\ and\ \citenamefont {Li}}]{han2016spectral}%
  \BibitemOpen
  \bibfield  {author} {\bibinfo {author} {\bibfnamefont {J.}~\bibnamefont
  {Han}}, \bibinfo {author} {\bibfnamefont {T.}~\bibnamefont {Vogt}}, \ and\
  \bibinfo {author} {\bibfnamefont {W.}~\bibnamefont {Li}},\ }\bibfield
  {title} {\enquote {\bibinfo {title} {Spectral shift and dephasing of
  electromagnetically induced transparency in an interacting rydberg gas},}\
  }\href@noop {} {\bibfield  {journal} {\bibinfo  {journal} {Physical Review
  A}\ }\textbf {\bibinfo {volume} {94}},\ \bibinfo {pages} {043806} (\bibinfo
  {year} {2016})}\BibitemShut {NoStop}%
\bibitem [{\citenamefont {Krauth}(2006)}]{krauth2006statistical}%
  \BibitemOpen
  \bibfield  {author} {\bibinfo {author} {\bibfnamefont {W.}~\bibnamefont
  {Krauth}},\ }\href@noop {} {\emph {\bibinfo {title} {Statistical mechanics :
  algorithms and computations}}},\ Vol.~\bibinfo {volume} {13}\ (\bibinfo
  {publisher} {OUP Oxford},\ \bibinfo {year} {2006})\BibitemShut {NoStop}%
\bibitem [{\citenamefont {Singer}\ \emph {et~al.}(2005)\citenamefont {Singer},
  \citenamefont {Stanojevic}, \citenamefont {Weidem{\"u}ller},\ and\
  \citenamefont {C{\^o}t{\'e}}}]{singer2005long}%
  \BibitemOpen
  \bibfield  {author} {\bibinfo {author} {\bibfnamefont {K.}~\bibnamefont
  {Singer}}, \bibinfo {author} {\bibfnamefont {J.}~\bibnamefont {Stanojevic}},
  \bibinfo {author} {\bibfnamefont {M.}~\bibnamefont {Weidem{\"u}ller}}, \ and\
  \bibinfo {author} {\bibfnamefont {R.}~\bibnamefont {C{\^o}t{\'e}}},\
  }\bibfield  {title} {\enquote {\bibinfo {title} {Long-range interactions
  between alkali rydberg atom pairs correlated to the ns--ns, np--np and nd--nd
  asymptotes},}\ }\href@noop {} {\bibfield  {journal} {\bibinfo  {journal}
  {Journal of Physics B: Atomic, Molecular and Optical Physics}\ }\textbf
  {\bibinfo {volume} {38}},\ \bibinfo {pages} {S295} (\bibinfo {year}
  {2005})}\BibitemShut {NoStop}%
\bibitem [{\citenamefont {G{\"a}rttner}\ and\ \citenamefont
  {Evers}(2013)}]{garttner2013nonlinear}%
  \BibitemOpen
  \bibfield  {author} {\bibinfo {author} {\bibfnamefont {M.}~\bibnamefont
  {G{\"a}rttner}}\ and\ \bibinfo {author} {\bibfnamefont {J.}~\bibnamefont
  {Evers}},\ }\bibfield  {title} {\enquote {\bibinfo {title} {Nonlinear
  absorption and density-dependent dephasing in rydberg
  electromagnetically-induced-transparency media},}\ }\href@noop {} {\bibfield
  {journal} {\bibinfo  {journal} {Phys. Rev. A}\ }\textbf {\bibinfo {volume}
  {88}},\ \bibinfo {pages} {033417} (\bibinfo {year} {2013})}\BibitemShut
  {NoStop}%
\bibitem [{\citenamefont {Schempp}\ \emph {et~al.}(2014)\citenamefont
  {Schempp}, \citenamefont {G{\"u}nter}, \citenamefont {Robert-de
  Saint-Vincent}, \citenamefont {Hofmann}, \citenamefont {Breyel},
  \citenamefont {Komnik}, \citenamefont {Sch{\"o}nleber}, \citenamefont
  {G{\"a}rttner}, \citenamefont {Evers}, \citenamefont {Whitlock} \emph
  {et~al.}}]{schempp2014full}%
  \BibitemOpen
  \bibfield  {author} {\bibinfo {author} {\bibfnamefont {H.}~\bibnamefont
  {Schempp}}, \bibinfo {author} {\bibfnamefont {G.}~\bibnamefont {G{\"u}nter}},
  \bibinfo {author} {\bibfnamefont {M.}~\bibnamefont {Robert-de
  Saint-Vincent}}, \bibinfo {author} {\bibfnamefont {C.~S.}\ \bibnamefont
  {Hofmann}}, \bibinfo {author} {\bibfnamefont {D.}~\bibnamefont {Breyel}},
  \bibinfo {author} {\bibfnamefont {A.}~\bibnamefont {Komnik}}, \bibinfo
  {author} {\bibfnamefont {DW}~\bibnamefont {Sch{\"o}nleber}}, \bibinfo
  {author} {\bibfnamefont {Martin}\ \bibnamefont {G{\"a}rttner}}, \bibinfo
  {author} {\bibfnamefont {J{\"o}rg}\ \bibnamefont {Evers}}, \bibinfo {author}
  {\bibfnamefont {S}~\bibnamefont {Whitlock}},  \emph {et~al.},\ }\bibfield
  {title} {\enquote {\bibinfo {title} {Full counting statistics of laser
  excited rydberg aggregates in a one-dimensional geometry},}\ }\href@noop {}
  {\bibfield  {journal} {\bibinfo  {journal} {Phys. Rev. Lett.}\ }\textbf
  {\bibinfo {volume} {112}},\ \bibinfo {pages} {013002} (\bibinfo {year}
  {2014})}\BibitemShut {NoStop}%
\bibitem [{\citenamefont {Valado}\ \emph {et~al.}(2015)\citenamefont {Valado},
  \citenamefont {Simonelli}, \citenamefont {Hoogerland}, \citenamefont
  {Lesanovsky}, \citenamefont {Garrahan}, \citenamefont {Arimondo},
  \citenamefont {Ciampini},\ and\ \citenamefont
  {Morsch}}]{valado2015experimental}%
  \BibitemOpen
  \bibfield  {author} {\bibinfo {author} {\bibfnamefont {M.~M.}\ \bibnamefont
  {Valado}}, \bibinfo {author} {\bibfnamefont {C.}~\bibnamefont {Simonelli}},
  \bibinfo {author} {\bibfnamefont {M.~D.}\ \bibnamefont {Hoogerland}},
  \bibinfo {author} {\bibfnamefont {I.}~\bibnamefont {Lesanovsky}}, \bibinfo
  {author} {\bibfnamefont {J.~P.}\ \bibnamefont {Garrahan}}, \bibinfo {author}
  {\bibfnamefont {E.}~\bibnamefont {Arimondo}}, \bibinfo {author}
  {\bibfnamefont {D.}~\bibnamefont {Ciampini}}, \ and\ \bibinfo {author}
  {\bibfnamefont {O.}~\bibnamefont {Morsch}},\ }\bibfield  {title} {\enquote
  {\bibinfo {title} {Experimental observation of controllable kinetic
  constraints in a cold atomic gas},}\ }\href@noop {} {\bibfield  {journal}
  {\bibinfo  {journal} {arXiv preprint arXiv:1508.04384}\ } (\bibinfo {year}
  {2015})}\BibitemShut {NoStop}%
\bibitem [{\citenamefont {Pritchard}\ \emph {et~al.}(2010)\citenamefont
  {Pritchard}, \citenamefont {Maxwell}, \citenamefont {Gauguet}, \citenamefont
  {Weatherill}, \citenamefont {Jones},\ and\ \citenamefont
  {Adams}}]{pritchard2010cooperative}%
  \BibitemOpen
  \bibfield  {author} {\bibinfo {author} {\bibfnamefont {J.~D.}\ \bibnamefont
  {Pritchard}}, \bibinfo {author} {\bibfnamefont {D.}~\bibnamefont {Maxwell}},
  \bibinfo {author} {\bibfnamefont {Al.}\ \bibnamefont {Gauguet}}, \bibinfo
  {author} {\bibfnamefont {K.~J.}\ \bibnamefont {Weatherill}}, \bibinfo
  {author} {\bibfnamefont {M.~P.~A.}\ \bibnamefont {Jones}}, \ and\ \bibinfo
  {author} {\bibfnamefont {C.~S.}\ \bibnamefont {Adams}},\ }\bibfield  {title}
  {\enquote {\bibinfo {title} {Cooperative atom-light interaction in a
  blockaded rydberg ensemble},}\ }\href@noop {} {\bibfield  {journal} {\bibinfo
   {journal} {Phys. Rev. Lett.}\ }\textbf {\bibinfo {volume} {105}},\ \bibinfo
  {pages} {193603} (\bibinfo {year} {2010})}\BibitemShut {NoStop}%
\bibitem [{\citenamefont {Weimer}\ \emph {et~al.}(2008)\citenamefont {Weimer},
  \citenamefont {L{\"o}w}, \citenamefont {Pfau},\ and\ \citenamefont
  {B{\"u}chler}}]{weimer2008quantum}%
  \BibitemOpen
  \bibfield  {author} {\bibinfo {author} {\bibfnamefont {H.}~\bibnamefont
  {Weimer}}, \bibinfo {author} {\bibfnamefont {R.}~\bibnamefont {L{\"o}w}},
  \bibinfo {author} {\bibfnamefont {T.}~\bibnamefont {Pfau}}, \ and\ \bibinfo
  {author} {\bibfnamefont {H.~P.}\ \bibnamefont {B{\"u}chler}},\ }\bibfield
  {title} {\enquote {\bibinfo {title} {Quantum critical behavior in strongly
  interacting rydberg gases},}\ }\href@noop {} {\bibfield  {journal} {\bibinfo
  {journal} {Phys. Rev. Lett.}\ }\textbf {\bibinfo {volume} {101}},\ \bibinfo
  {pages} {250601} (\bibinfo {year} {2008})}\BibitemShut {NoStop}%
\bibitem [{\citenamefont {Petrosyan}\ \emph {et~al.}(2011)\citenamefont
  {Petrosyan}, \citenamefont {Otterbach},\ and\ \citenamefont
  {Fleischhauer}}]{petrosyan2011electromagnetically}%
  \BibitemOpen
  \bibfield  {author} {\bibinfo {author} {\bibfnamefont {D.}~\bibnamefont
  {Petrosyan}}, \bibinfo {author} {\bibfnamefont {J.}~\bibnamefont
  {Otterbach}}, \ and\ \bibinfo {author} {\bibfnamefont {M.}~\bibnamefont
  {Fleischhauer}},\ }\bibfield  {title} {\enquote {\bibinfo {title}
  {Electromagnetically induced transparency with rydberg atoms},}\ }\href@noop
  {} {\bibfield  {journal} {\bibinfo  {journal} {Phys. Rev. Lett.}\ }\textbf
  {\bibinfo {volume} {107}},\ \bibinfo {pages} {213601} (\bibinfo {year}
  {2011})}\BibitemShut {NoStop}%
\bibitem [{\citenamefont {Sevin{\c{c}}li}\ \emph
  {et~al.}(2011{\natexlab{a}})\citenamefont {Sevin{\c{c}}li}, \citenamefont
  {Henkel}, \citenamefont {Ates},\ and\ \citenamefont
  {Pohl}}]{sevinccli2011nonlocal}%
  \BibitemOpen
  \bibfield  {author} {\bibinfo {author} {\bibfnamefont {S.}~\bibnamefont
  {Sevin{\c{c}}li}}, \bibinfo {author} {\bibfnamefont {N.}~\bibnamefont
  {Henkel}}, \bibinfo {author} {\bibfnamefont {C.}~\bibnamefont {Ates}}, \ and\
  \bibinfo {author} {\bibfnamefont {T.}~\bibnamefont {Pohl}},\ }\bibfield
  {title} {\enquote {\bibinfo {title} {Nonlocal nonlinear optics in cold
  rydberg gases},}\ }\href@noop {} {\bibfield  {journal} {\bibinfo  {journal}
  {Phys. Rev. Lett.}\ }\textbf {\bibinfo {volume} {107}},\ \bibinfo {pages}
  {153001} (\bibinfo {year} {2011}{\natexlab{a}})}\BibitemShut {NoStop}%
\bibitem [{\citenamefont {Sevin{\c{c}}li}\ \emph
  {et~al.}(2011{\natexlab{b}})\citenamefont {Sevin{\c{c}}li}, \citenamefont
  {Ates}, \citenamefont {Pohl}, \citenamefont {Schempp}, \citenamefont
  {Hofmann}, \citenamefont {G{\"u}nter}, \citenamefont {Amthor}, \citenamefont
  {Weidem{\"u}ller}, \citenamefont {Pritchard}, \citenamefont {Maxwell} \emph
  {et~al.}}]{sevinccli2011quantum}%
  \BibitemOpen
  \bibfield  {author} {\bibinfo {author} {\bibfnamefont {S.}~\bibnamefont
  {Sevin{\c{c}}li}}, \bibinfo {author} {\bibfnamefont {C.}~\bibnamefont
  {Ates}}, \bibinfo {author} {\bibfnamefont {T.}~\bibnamefont {Pohl}}, \bibinfo
  {author} {\bibfnamefont {H.}~\bibnamefont {Schempp}}, \bibinfo {author}
  {\bibfnamefont {C.~S.}\ \bibnamefont {Hofmann}}, \bibinfo {author}
  {\bibfnamefont {G.}~\bibnamefont {G{\"u}nter}}, \bibinfo {author}
  {\bibfnamefont {T.}~\bibnamefont {Amthor}}, \bibinfo {author} {\bibfnamefont
  {M.}~\bibnamefont {Weidem{\"u}ller}}, \bibinfo {author} {\bibfnamefont
  {J.~D.}\ \bibnamefont {Pritchard}}, \bibinfo {author} {\bibfnamefont
  {D.}~\bibnamefont {Maxwell}},  \emph {et~al.},\ }\bibfield  {title} {\enquote
  {\bibinfo {title} {Quantum interference in interacting three-level rydberg
  gases: coherent population trapping and electromagnetically induced
  transparency},}\ }\href@noop {} {\bibfield  {journal} {\bibinfo  {journal}
  {Journal of Physics B: Atomic, Molecular and Optical Physics}\ }\textbf
  {\bibinfo {volume} {44}},\ \bibinfo {pages} {184018} (\bibinfo {year}
  {2011}{\natexlab{b}})}\BibitemShut {NoStop}%
\bibitem [{\citenamefont {DeSalvo}\ \emph {et~al.}(2016)\citenamefont
  {DeSalvo}, \citenamefont {Aman}, \citenamefont {Gaul}, \citenamefont {Pohl},
  \citenamefont {Yoshida}, \citenamefont {Burgd{\"o}rfer}, \citenamefont
  {Hazzard}, \citenamefont {Dunning},\ and\ \citenamefont
  {Killian}}]{desalvo2016rydberg}%
  \BibitemOpen
  \bibfield  {author} {\bibinfo {author} {\bibfnamefont {B.~J.}\ \bibnamefont
  {DeSalvo}}, \bibinfo {author} {\bibfnamefont {J.~A.}\ \bibnamefont {Aman}},
  \bibinfo {author} {\bibfnamefont {C.}~\bibnamefont {Gaul}}, \bibinfo {author}
  {\bibfnamefont {T.}~\bibnamefont {Pohl}}, \bibinfo {author} {\bibfnamefont
  {S.}~\bibnamefont {Yoshida}}, \bibinfo {author} {\bibfnamefont
  {J.}~\bibnamefont {Burgd{\"o}rfer}}, \bibinfo {author} {\bibfnamefont
  {K.~R.~A.}\ \bibnamefont {Hazzard}}, \bibinfo {author} {\bibfnamefont
  {F.~B.}\ \bibnamefont {Dunning}}, \ and\ \bibinfo {author} {\bibfnamefont
  {T.~C.}~\bibnamefont {Killian}},\ }\bibfield  {title} {\enquote {\bibinfo
  {title} {Rydberg-blockade effects in autler-townes spectra of ultracold
  strontium},}\ }\href@noop {} {\bibfield  {journal} {\bibinfo  {journal}
  {Physical Review A}\ }\textbf {\bibinfo {volume} {93}},\ \bibinfo {pages}
  {022709} (\bibinfo {year} {2016})}\BibitemShut {NoStop}%
\bibitem [{\citenamefont {Han}\ \emph {et~al.}(2015)\citenamefont {Han},
  \citenamefont {Vogt}, \citenamefont {Manjappa}, \citenamefont {Guo},
  \citenamefont {Kiffner},\ and\ \citenamefont {Li}}]{han2015lensing}%
  \BibitemOpen
  \bibfield  {author} {\bibinfo {author} {\bibfnamefont {J.}~\bibnamefont
  {Han}}, \bibinfo {author} {\bibfnamefont {T.}~\bibnamefont {Vogt}}, \bibinfo
  {author} {\bibfnamefont {M.}~\bibnamefont {Manjappa}}, \bibinfo {author}
  {\bibfnamefont {R.}~\bibnamefont {Guo}}, \bibinfo {author} {\bibfnamefont
  {M.}~\bibnamefont {Kiffner}}, \ and\ \bibinfo {author} {\bibfnamefont
  {W.}~\bibnamefont {Li}},\ }\bibfield  {title} {\enquote {\bibinfo {title}
  {Lensing effect of electromagnetically induced transparency involving a
  rydberg state},}\ }\href@noop {} {\bibfield  {journal} {\bibinfo  {journal}
  {Phys. Rev. A}\ }\textbf {\bibinfo {volume} {92}},\ \bibinfo {pages} {063824}
  (\bibinfo {year} {2015})}\BibitemShut {NoStop}%
\bibitem [{\citenamefont {G{\'o}rska}\ and\ \citenamefont
  {Penson}(2012)}]{gorska2012levy}%
  \BibitemOpen
  \bibfield  {author} {\bibinfo {author} {\bibfnamefont {K.}~\bibnamefont
  {G{\'o}rska}}\ and\ \bibinfo {author} {\bibfnamefont {K.~A.}\ \bibnamefont
  {Penson}},\ }\bibfield  {title} {\enquote {\bibinfo {title} {L{\'e}vy stable
  distributions via associated integral transform},}\ }\href@noop {} {\bibfield
   {journal} {\bibinfo  {journal} {Journal of Mathematical Physics}\ }\textbf
  {\bibinfo {volume} {53}},\ \bibinfo {pages} {053302} (\bibinfo {year}
  {2012})}\BibitemShut {NoStop}%
\bibitem [{\citenamefont {Mayer}\ and\ \citenamefont
  {Montroll}(1941)}]{mayer1941molecular}%
  \BibitemOpen
  \bibfield  {author} {\bibinfo {author} {\bibfnamefont {J.~E.}\ \bibnamefont
  {Mayer}}\ and\ \bibinfo {author} {\bibfnamefont {E.}~\bibnamefont
  {Montroll}},\ }\bibfield  {title} {\enquote {\bibinfo {title} {Molecular
  distribution},}\ }\href@noop {} {\bibfield  {journal} {\bibinfo  {journal}
  {The Journal of Chemical Physics}\ }\textbf {\bibinfo {volume} {9}},\
  \bibinfo {pages} {2--16} (\bibinfo {year} {1941})}\BibitemShut {NoStop}%
\end{thebibliography}
\end{document}